\documentclass[sigconf]{acmart}
\usepackage{tcolorbox}
\usepackage{enumitem}
\usepackage{subcaption}

\def\N{18}
\def\vf{\textsc{Value Faces}}

\AtBeginDocument{%
  }

\setcopyright{none}

\begin{document}

\title[\textbf{\vf{}}: Surfacing How Self-Presentation Shifts Across Relationships]{ \textbf{\vf{}}: \\Surfacing How Self-Presentation Shifts Across Relationships}


\author{Gabriel Koo}
\affiliation{%
  \institution{University of Michigan}
  \city{Ann Arbor}
  \state{Michigan}
  \country{USA}
}
\email{gabekoo@umich.edu}

\author{Rayhan Rashed}
\affiliation{%
  \institution{University of Michigan}
  \city{Ann Arbor}
  \state{Michigan}
  \country{USA}
}
\email{rayrash@umich.edu}

\author{Farnaz Jahanbakhsh}
\affiliation{%
  \institution{University of Michigan}
  \city{Ann Arbor}
  \state{Michigan}
  \country{USA}
}
\email{farnaz@umich.edu}

\renewcommand{\shortauthors}{Koo, Rashed, and Jahanbakhsh}

\newcommand{\gabriel}[1]{\textbf{\sffamily{\textcolor{dodgerblue}{[#1 -- Gabriel]}}}}
\newcommand{\farnaz}[1]{\textbf{\sffamily{\textcolor{red}{[#1 -- Farnaz]}}}}
\newcommand{\rayhan}[1]{\textbf{\sffamily{\textcolor{violet}{[#1 -- Rayhan]}}}}

\begin{abstract}


\noindent People present different aspects of themselves across relationships. Computational work has captured such variation in communication style. But this variation also extends to which principles people foreground or background in a particular relationship---i.e., in the values they express and how they balance them. We conceptualize these relationship-specific expressions of values as \emph{demonstrated values}. To make demonstrated values visible, we introduce \vf{}, a system that analyzes a person's existing chat histories from their everyday messaging platforms using Schwartz's ten basic human values and produces separate value profiles for their different relationships. In a mixed-methods study(N=18), we find that the resulting value profiles distinguished participants' relational contexts with twice the odds of guessing, while system-inferred differences across relationships aligned with participants' perceptions of those differences. Participants used these profiles to articulate previously implicit differences in how they presented themselves, connect them to roles and changes over time, and reconsider their self-assessments.


\end{abstract}



\begin{teaserfigure}
    \includegraphics[width=\textwidth]{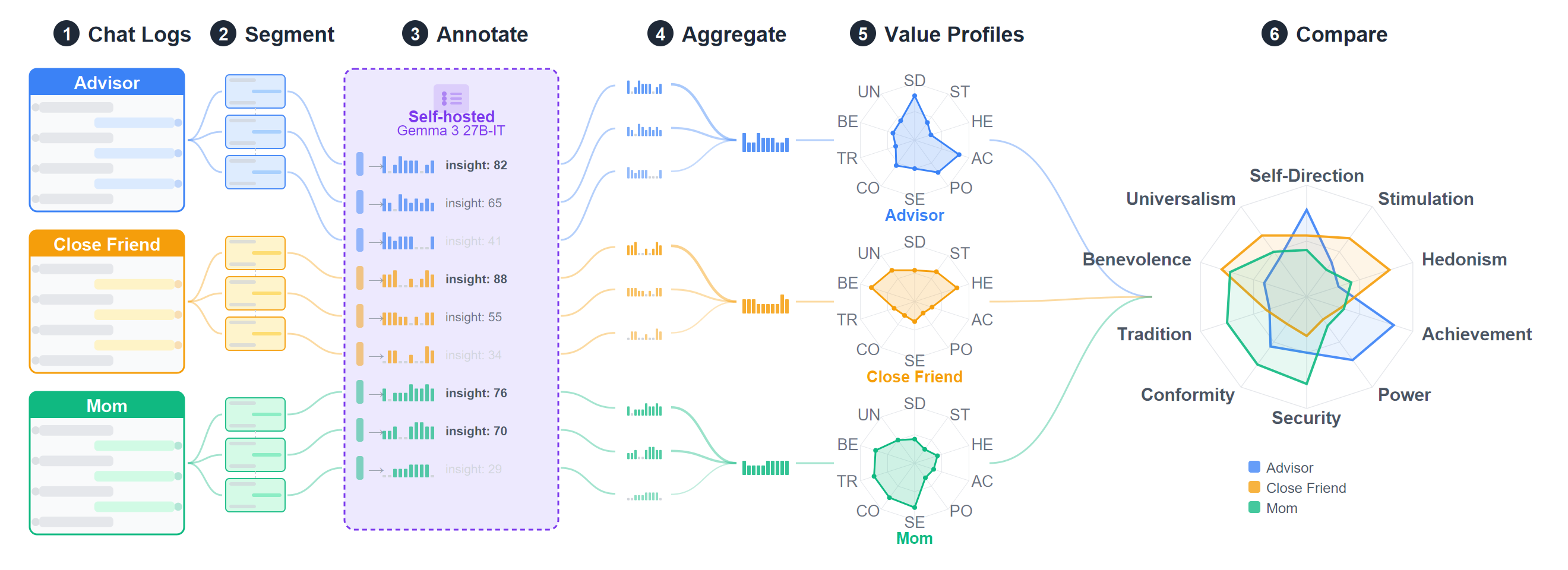}
    \caption{
        The Value Faces pipeline. Chat logs from multiple relational contexts (1) are segmented into conversation units (2), annotated by a locally-hosted LLM for ten Schwartz values (3), aggregated using insight and confidence weighting (4), producing per-context value profiles (5) that can be compared quantitatively across relationships (6).
    }
    \label{fig:pipeline}
    \Description{Six-stage pipeline diagram showing the transformation of chat logs into comparable value profiles. Stages flow left to right, connected by color-coded flow lines. Stage 1 (Chat Logs) shows three labeled chat log blocks for Advisor, Close Friend, and Mom, each with its own chat transcript. Stage 2 (Segment) shows these messages divided into discrete, manageable conversation units. Stage 3 (Annotate) shows our locally-hosted model producing bar chart annotations for each conversation unit with insightfulness scores. Stage 4 (Aggregate) shows bar charts being combined with insight and confidence weighting. Stage 5 (Value Profiles) shows three separate radar charts, one per relational context, each plotting ten Schwartz values as axes. The radar charts differ in shape: the Advisor profile skews toward Self-Direction and Achievement, the Close Friend profile is more balanced, and the Mom profile emphasizes Conformity and Benevolence. Stage 6 (Compare) overlays all three radar charts with a shared legend, enabling cross-context comparison.}
\end{teaserfigure}


\maketitle


\vspace{-3pt}
\section{Introduction}

People present themselves differently to the different people in their lives. Someone can argue for her own judgment with a colleague and defer to her mother's on the same question, and she may sense the difference without being able to say what it consists of. Today, these relational dynamics increasingly unfold across messaging platforms, leaving rich digital traces of how a person communicates across different spheres of her life. Inference over the digital traces people leave is by now routine, and it runs toward ends set by platforms such as engagement~\cite{zuboff_surveillance}, ad targeting~\cite{ur2012smart}, and retention~\cite{budak2017threading, kilimci2020sentiment}. We envision instead turning these traces toward the people producing them, making aspects of themselves that are otherwise hard to see available for their own self-reflection. Here, we pursue this vision by asking what people's digital traces can reveal about how they present themselves across their relationships.

Social science has long recognized these relational `fronts'. 
Sociologist Erving Goffman famously treats social interaction as performance, where people select different fronts for different audiences and adjust each to the demands of a particular relationship~\cite{goffman1959presentation}. Computational work has examined such adjustments through linguistic markers such as word choice, politeness, and style more broadly \cite{pavalanathan2015audience, danescu2013computational}. 
While these markers capture the mechanics of self-presentation, they do not capture the priorities expressed through those adjustments. Making these priorities explicit can support reflection not just on how a person communicates across relationships, but on what she appears to prioritize in each.

\emph{Human values} provide a common vocabulary for describing these priorities. 
Values are core motivations that individuals hold in common but prioritize differently~\cite{schwartz1992universals}.
To surface the values behind an individual's relational fronts, we adopt a well-established value framework in cultural psychology, Schwartz's Theory of Basic Human Values~\cite{schwartz1992universals,schwartz2012overview}. The theory identifies ten broad motivational goals and organizes them in a circumplex based on their compatibility and conflict. But using values to understand relational variation raises the question of how to characterize the values a person expresses within a particular relationship rather than her values in general.
The standard instrument for measuring these values, the Portrait Values Questionnaire (PVQ), does the latter. It asks individuals to compare themselves with brief portraits and summarizes their responses in a single profile of their general value priorities~\cite{schwartz2001extending}. Because this profile characterizes the person as a whole, it cannot show how value expression differs across relationships.

To make this relational variation visible, we introduce the construct of \textbf{\textit{demonstrated values}}, the values a person expresses within a particular relational context. Relationships differ in what they call for---which topics arise, which positions a person finds herself defending---so they give different values occasion to surface. 
Unlike the general value profile produced by the PVQ, a profile of demonstrated values characterizes a single relational context. No relational profile is treated as a more ``authentic'' account of the person than another. Each reflects how the person navigates that relationship and the variation among profiles is what we measure and make visible.

We infer an individual's demonstrated values from her exchanges with each of the people in her life, and present them back to her in a form she can inspect and compare.
We do this through \textbf{\vf{}}, a privacy-first web application that derives demonstrated value profiles from chat histories exported from everyday messaging platforms including Slack, Discord, and WhatsApp. Each profile represents one relational context and draws on conversations accumulated over time with the same interlocutor or group. To construct these profiles, \vf{} segments each chat log into conversation units and annotates them for Schwartz's ten values using a locally hosted LLM classifier. Because these units vary in how much they reveal about a person's values, 
those with stronger value evidence receive greater weight during aggregation. The resulting profiles feed into an interactive dashboard where users can compare their demonstrated values across relationships, and over time, and inspect the conversation units contributing to each profile.  


We evaluated \vf{} through a mixed-methods evaluation with \N{} participants who contributed 79 chat logs in total. We first tested whether demonstrated values distinguish relationships, that is, whether a profile built from some of a user's conversation units with an interlocutor can identify which relationship a held-out conversation came from. Using only the inferred value profiles, a classifier assigned held-out conversation units to their correct interlocutor at twice the odds expected by chance $(\text{OR} = 2.00, p < .001)$. Structure alone, however, could be an artifact of the pipeline, so we compared the system's inferences against what participants report about their own relationships. Participants recorded the value differences they perceived across relationships \textit{\textbf{before}} seeing any output, and those reports correspond significantly with the differences \vf{} infers $(\text{OR} = 1.78, p < .001)$. This validates the pipeline end to end. As participants navigated the dashboard, we asked where the profiles matched their sense of each relationship, where they did not, and what accounted for the differences. Inspecting their profiles, participants recognized patterns in their self-presentation they had sensed but never articulated, traced value shifts to life events and role changes within relationships, and used the value vocabulary to revise their own self-assessments.

\noindent In summary, this paper contributes the following:
\begin{enumerate}
    \item The construct of \textbf{demonstrated values}, which characterizes the values a person expresses within a particular relational context and gives cross-relationship variation in self-presentation a form that can be measured and compared.
    \item \textbf{\vf{}}, a system that infers demonstrated values from multi-platform chat logs and visualizes how they vary across a person's relationships, with every profile linked to the conversation units behind it.  
    \item {Empirical evidence} from a mixed-methods study with \N{} participants whose demonstrated value profiles produced by \vf{} vary systematically across relationships, match what participants perceive about their own relationships, and give people a vocabulary for reflecting on how they come across.
\end{enumerate}

\section{Related Work}

\vf{} connects three lines of research: the psychology of human values and how they are measured, self-presentation and audience management across relational contexts, and the design of systems that help people reflect on their own data.
\subsection{Human Values and Their Measurement}

Human values are enduring motivational principles that guide judgments, priorities, and behavior across situations~\cite{schwartz1992universals}. Several frameworks exist for studying human values, including Schwartz's Theory of Basic Human Values~\cite{schwartz1992universals}, Rokeach's terminal and instrumental values~\cite{rokeach1973nature}, the World Values Survey~\cite{inglehart2005modernization}, Hofstede's cultural dimensions~\cite{hofstede2001culture}, and Moral Foundations Theory~\cite{graham2013moral}. 
In this work, we draw on Schwartz's Theory of Basic Human Values for two reasons. First, it frames values as individual priorities that vary from person to person, rather than as culture-level dimensions. Second, it arranges ten motivational value types---Self-Direction, Stimulation, Hedonism, Achievement, Power, Security, Conformity, Tradition, Benevolence, and Universalism---in a circumplex where adjacent values share motivational underpinnings and opposing values express competing goals~\cite{schwartz1992universals, schwartz2012overview}. That structure makes value trade-offs interpretable.

The standard instrument for measuring Schwartz values, the Portrait Values Questionnaire (PVQ) and its revised form PVQ-RR~\cite{schwartz2001extending, schwartz2003proposal, schwartz2022measuring}, asks people how much they resemble a set of short portraits and produces a single ranking of an individual's value priorities through that self-report. Values in Schwartz's formulation transcend specific actions and situations~\cite{schwartz1992universals}. Therefore, the PVQ recovers a person's general motivational orientation and orientations measured this way are relatively stable over time~\cite{bardi2003values}. Nothing in this instrument, however, speaks to whether the priorities a person \emph{expresses} differ depending on who they are talking to. Asking this question is consistent with Schwartz's own account that values influence action when they are relevant in the context and hence likely to be activated~\cite{schwartz1992universals}. A relational context can thus be understood as activating certain values, shaping which ones become expressed in behavior and observable to others. Our construct of \emph{demonstrated values} captures this phenomenon, and rather than recovering a person's general value orientation, asks what values a particular relational context brings to the surface.

\paragraph{Values in text.}
Labeling human values in text has been of interest to HCI and NLP communities for a range of applications including auditing feed algorithms~\cite{epstein2026value}, aligning social media feeds with values~\cite{jahanbakhsh2025value,epstein2025measuring,jia2024embedding}, and aligning generative AI on certain human values~\cite{jiang2025can,sorensen2025value}. Method-wise, labeling of values in text has progressed from lexicon-based approaches~\cite{ponizovskiy2020development} to model-based classification, including the SemEval-2023 ValueEval, which identifies values in argumentative text~\cite{kiesel2023semeval} and LLM-based classification of value expressions~\cite{epstein2025measuring,epstein2026value}. Because value definitions from the social sciences are precise constructs, LLMs prompted with these definitions produce value labels that align closely with human annotations~\cite{jia2024embedding, jahanbakhsh2025value, kolluri2025alexandria}. 
Recent work further extends value inference from isolated posts to multi-turn conversations. For instance, multiple exchanges around a post on Reddit have been used to build profiles of users' expressed values~\cite{vuyyuru2026persuasion}. Similarly, month-long interactions with an LLM-powered chatbot can serve as the basis for constructing detailed user value profiles~\cite{yun2026vapt}.

Across this body of work, value detection recovers a single profile per person, whether from isolated posts or from an extended conversation. \vf{} operates on conversations people have already had, on the platforms they already use, and produces a separate profile for each relationship so those differences become visible and comparable.

\vspace{-4pt}
\subsection{Self-Presentation \& Audience Management}
Erving Goffman conceptualizes social interaction as performance. In his dramaturgical model of self-presentation, he theorizes that individuals select different ``fronts'' for different audiences. The dynamic self emerging through this is a dramatic effect of the scene, where audience, role, and setting operate as one interactional whole~\cite{goffman1959presentation}. Work on impression management has since formalized both the strategic and habitual dimensions of this process~\cite{leary1990impression, schlenker1980impression}. Communication Accommodation Theory further describes how such audience-sensitive adjustment can manifest in language, with speakers shifting linguistic style toward or away from interlocutors depending on relational goals~\cite{giles1991accommodation}. Across these frameworks, self-presentation is consistently described as an ongoing calibration to the perceived audience.

Walther's hyperpersonal model argues that the asynchronous, editable nature of computer-mediated communication gives users greater control over self-presentation than face-to-face interaction, so the performances that result are more deliberately tailored to the audience~\cite{walther1996cmc}. Litt formalizes this through the concept of the \emph{imagined audience}, the mental model a communicator constructs of whom they are addressing~\cite{litt2012imagined, litt2016imagined}. 


However, this audience management leaves ample measurable traces in digital communications.
Computational studies have documented such traces in systematic shifts in formality, politeness, and lexical choice across platforms and audiences~\cite{pavalanathan2015audience, danescu2013computational}. 
All this work, however, operates at the level of linguistic markers. While meaningful on their own, these markers could be thought of as the \textit{mechanics} of self-presentation rather than
the \textit{priorities} and \textit{tradeoffs} expressed through it. Our system, \vf{} instead characterizes relational variation through the values a person foregrounds across relationships. In doing so, we use Schwartz's theory of human values to characterize a dimension of the relational self-presentation described by Goffman's dramaturgical framework.

\vspace{-4pt}
\subsection{Personal Informatics}

Personal informatics research designs systems that help people collect and reflect on their own data~\cite{li2010stage}. Most systems in this space track behavioral or physiological signals such as steps, sleep, screen time, or spending~\cite{li2011understanding, choe2014understanding}. More recent work has challenged the assumption that self-tracking needs to be goal-directed or tied to behavior change. Tracking practices are often exploratory or documentary~\cite{rooksby2014personal,epstein2015lived}, and data collected for one purpose can acquire reflective value over time as a record of one's past~\cite{elsden2016quantified}. A separate thread has pushed personal informatics beyond the individual, arguing that health management and self-understanding are inherently social and depend on a person's relationships and broader social context~\cite{murnane2018interpersonal, elsden2017quantified_social}. These expansions have broadened what personal informatics can be used for and whose data matters. The representations produced in this body of work, however, remain largely behavioral: counts, durations, and frequencies.

\vf{} builds on this understanding of personal informatics by operating on conversational data that people have already accumulated across their messaging platforms, requires no active tracking, and produces value profiles rather than behavioral summaries. These profiles make relational variation in value expression available for comparison and reflection across relationships.


\begin{figure*}[!t]
  \centering
  \includegraphics[width=\textwidth]{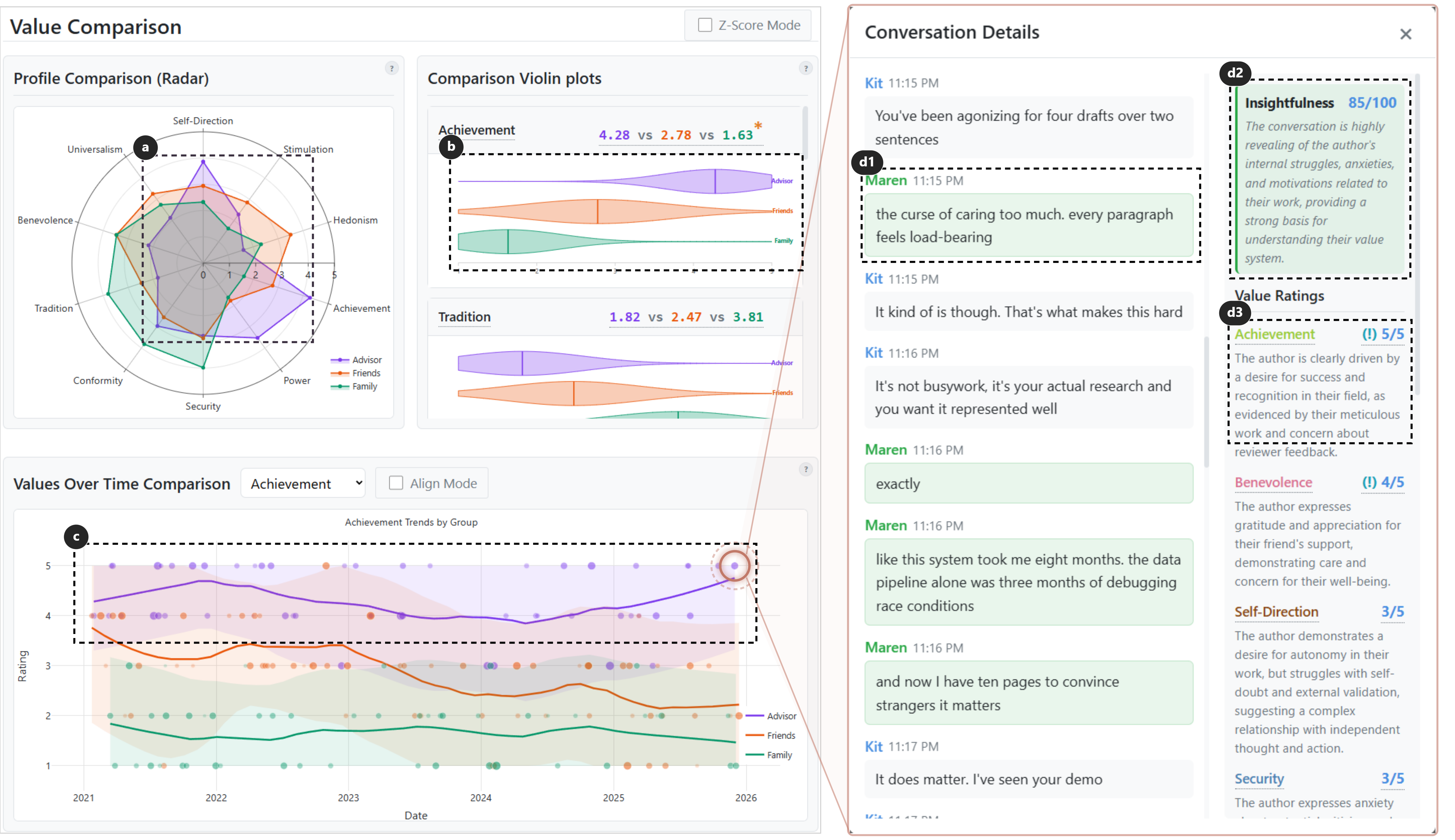}
  \caption{%
    System's cross-comparison of chat logs from friends, family, and an advisor, with focus on the Advisor context. The \textbf{(a) Radar chart} highlights the values the user demonstrates above their norm with their Advisor: Self-Direction, Achievement, and Power. 
    \textbf{(b) Violin plots} are sorted by inter-context variation. They reveal Achievement as the user's most context-sensitive value, demonstrated most strongly in the Advisor context.
    Selecting this violin plot displays the corresponding \textbf{(c) Values Over Time Comparison}, showing demonstrated Achievement with the Advisor stayed consistently high, as opposed to decreasing over time like in the Friends context. Selecting any point brings up the \textbf{(d) Conversation Details} modal for that point, which includes (d1) the per-turn chat history, (d2) the insightfulness rating for that conversation, and (d3) the LLM's rating for that conversation for every relevant value.  
  }
  \Description{Value Faces dashboard comparing demonstrated values across Advisor, Friends, and Family contexts. (a) shows an overlaid radar chart with ten Schwartz values as axes. The Advisor profile extends furthest toward Self-Direction, Achievement, and Power. (b) shows horizontal violin plots comparing value ratings across contexts, with aggregate Achievement scores of 4.28 for Advisor, 2.78 for Friends, and 1.63 for Family. (c) shows conversation-level Achievement ratings and smoothed trend lines from 2021 to 2026. Advisor ratings remain high, Friends ratings decrease, and Family ratings remain low. A highlighted point connects to the Conversation Details modal on the right. (d1) shows individual chat messages, (d2) shows an insightfulness score of 85/100 with explanatory text, and (d3) shows value ratings with the language model's explanations, including Achievement rated 5/5.}
  \label{fig:cross_comparison}
\end{figure*}
\vspace{-4pt}

\section{ \vf{}: System Design}
\label{sec:system}

\vf{} is a web application that transforms raw chat exports into interactive value profiles to aid user reflection on how they present themselves differently across relationships. Users upload chat logs from their existing messaging platforms. 
Each uploaded chat log constitutes a \textbf{relational context}: a one-on-one conversation partner or a group whose chat history the user exports as a single file. 
The system annotates each segment of the conversation for Schwartz's ten basic human values, aggregates the results into a value profile for each relational context, and infers higher-level clusters based on similarity in value space. The pipeline comprises four stages: (1)~client-side parsing of chat logs and conversation segmentation, (2)~LLM-based value annotation, (3)~weighted aggregation of values, and (4)~interactive visualization with cross-context comparison.

We describe each stage below in \S\ref{sec:segmentation}-\S\ref{sec:visualization}. Then, \S\ref{sec:evaluation} describes our study design and the statistical methods we use to test 1) whether people's value sets as inferred by our pipeline are distinguishable, and 2) to what degree the value differences inferred by our system align with participants' own perceived differences.

\subsection{Data Sourcing and Privacy}
\label{sec:privacy}
Because our pipeline necessarily processes private conversational text, we took several steps to limit exposure and retention of data.
First, raw chat files are parsed entirely within the user's browser. The server never receives the original export; only structured message records (author pseudonym, timestamp, message text) leave the client, and only after the user reviews a preview and explicitly confirms the upload. Second, before any text reaches the annotation model, all author identifiers are replaced with generic labels: the target user becomes \texttt{TARGET~AUTHOR} and all other participants become \texttt{CONVERSATION~PARTNER~$k$}. This helps prevent the model from conditioning on real names, profile metadata, or social context beyond what appears in the message text. Third, after annotation, we retain only de-identified numerical summaries (value ratings and aggregate statistics) in the research dataset. In accordance with our IRB protocol, we discard all message text after annotation. These same concerns motivated our choice to run the annotation model on infrastructure under our control (\S\ref{sec:annotation}) rather than transmitting private text to a commercial API.



Because people often use different platforms for different relation\-ships---for example, texting family on WhatsApp, coordinating with colleagues on Slack, or chatting with friends on Discord---cross-context comparison requires multi-platform support.
\vf{} accepts chat exports from five platforms: WhatsApp, Discord, Slack, iMessage, and Snapchat. We did not find any existing plugin to export chat from Discord, Slack, and Snapchat. Therefore, we wrote a browser extension plugin that facilitates chat export from those platforms.
Each platform-specific parser normalizes messages into a common \texttt{\{author, date, content\}} schema.
\vspace{-9pt}
\subsection{Conversation Segmentation}
\label{sec:segmentation}
Values are rarely expressed in a single utterance; they emerge across turns in an ongoing exchange. To capture this, we first partition each chat log into \textit{conversational units} and treat each unit as the basis for value labeling. We use a time-gap heuristic: messages are sorted chronologically and split whenever the interval between consecutive messages exceeds a certain threshold---in our study it is two hours.
In pilot testing with 10 chat logs across 4 platforms, we identified two failure modes: (1) Gaps below one hour frequently split topically continuous exchanges, 
breaking semantically continuous exchanges into fragments too short for meaningful value inference. (2) Above four hours, topically distinct conversations were merged such that conversations with different interaction premises and social goals were collapsed into a single unit, diluting the contextual specificity of the resulting value ratings. We found that two hours balanced these failure modes for the asynchronous messaging patterns typical for the platforms \vf{} supports. We did a sensitivity check on various time gaps, and the results are very consistent across gaps. More in Appendix \S\ref{sec:appendix_segmentation}.

We rely on insightfulness scores assigned during annotation (\S\ref{sec:annotation}) to down-weight units that carry little value signal. Brief and non-substantive exchanges such as greetings or unanswered availability checks still enter the dataset, but contribute negligibly to the aggregated profile due to their low insightfulness scores.
\vspace{-4pt}
\subsection{Value Annotation}
\label{sec:annotation}

We submit each conversation independently to an LLM for value annotation. The system prompt instructs the model to act as a psychology assistant specializing in Schwartz's Theory of Basic Human Values~\cite{schwartz1992universals, schwartz2012overview} and to focus exclusively on the \texttt{TARGET~AUTHOR}, though the transcript of the full conversational unit, including the interlocutor's messages, is provided to preserve the context needed for value inference. For each of the ten Schwartz values, the model provides a one-sentence analysis describing how (or whether) the target author expresses the value, and an integer rating on a 1--5 scale (Figure~\ref{fig:cross_comparison}, d3) where 1 indicates that the author evinces the inverse of the value and 5 indicates strong demonstration. Importantly, if the conversation contains \textit{no} evidence for or against a value, the model assigns $-1$ (``No data'') rather than extrapolating. This distinction prevents the model from conflating absence of evidence with active opposition, a separation consistent with prior value measurement work~\cite{schwartz2003proposal,vuyyuru2026persuasion}. In pilot testing, omitting this option led models to assign artificially low ratings to values that were simply not discussed.

Beyond the per-value ratings, the annotation produces \textit{two} forms of confidence metadata. The {first} is a conversation-level \textit{insightfulness score} (0--100), estimating how useful the conversation unit as a whole is for value inference (Figure~\ref{fig:cross_comparison}, d2). 
The {second} is a per-value \textit{confidence score} (0--100), estimating how clearly each value is demonstrated in the unit. 
Both scores feed into the aggregation step (\S\ref{sec:aggregation}), where they determine how heavily each conversation unit and each per-value rating contributes to the final profile.
The full prompt and conversation formatting are in Appendix \S\ref{sec:appendix_prompt}.

\vspace{2pt}
\noindent\textbf{Model and hosting.}
To keep participants' conversational data within a controlled environment, we host the annotation model on university-managed AWS infrastructure rather than using a commercial API. After evaluating 9 open-source candidate models (Appendix \S\ref{sec:appendix_model}) for annotation quality, JSON formatting reliability, and consistency across repeated runs, we selected Gemma~3-27B-IT.
We set temperature to 0 and a fixed random seed for reproducibility. Conversation units that exceed the model's context window are truncated using a middle-out strategy that preserves opening and closing turns while removing middle segments~\cite{schegloff1973opening}.

\vspace{-4pt}
\subsection{Aggregation}
\label{sec:aggregation}

As described in \S\ref{sec:segmentation}, we retain all conversation units regardless of length and instead control their influence at aggregation time. The weighting addresses two sources of heterogeneity: conversation-level insightfulness and per-value confidence score.

For a set of annotated conversational units from a given chat log, the aggregate rating for each value $k$ is computed as a weighted mean. Let $s_{ik}$ be the rating for value $k$ in unit $i$ (excluding units where $s_{ik} = -1$), let $\mathrm{ins}_i \in [0, 100]$ be the insightfulness score for unit $i$, let $\mathrm{conf}_{ik} \in [0, 100]$ be the confidence score for value $k$ in unit $i$. Also, let $\alpha$ be the insight weight parameter. The weight for unit $i$ and value $k$ is:
\vspace{-2pt}

\[
w_{ik} = \underbrace{\left[(1 - \alpha) + \alpha \cdot \frac{\mathrm{ins}_i}{100}\right]}_{\text{insight weight}} \cdot \underbrace{\frac{\mathrm{conf}_{ik}}{100}}_{\text{confidence}}
\]

We set $\alpha = 0.80$, based on our pilots. At this value, a maximally insightful conversation unit receives five times the weight of a minimally insightful one, which produced profiles that our pilot participants found more consistent with their own understanding of each relationship than lower $\alpha$ values did. 
The confidence component ensures that a value merely alluded to in passing contributes less than one discussed at length.

\noindent Aggregate rating for value 
$k$ with a given interlocutor is then:
\vspace{-2pt}
\[
\bar{s}_k = \frac{\sum_i w_{ik} \cdot s_{ik}}{\sum_i w_{ik}}
\]

We compute a 95\% margin of error using the weighted sample variance $\mathrm{Var}_k^{(w)}$ and Kish's effective sample size $n_{\mathrm{eff}}$ to account for unequal weights, with $z = 1.96$ as the critical value:
\vspace{-2pt}
\[
\mathrm{MoE}_k = z \cdot \frac{\sqrt{\mathrm{Var}_k^{(w)}}}{\sqrt{n_{\mathrm{eff}}}}
\]

We call the resulting profile $\{\bar{s}_k \pm \mathrm{MoE}_k\}$ a \textit{value face}: the demonstrated value profile a person projects in a specific relational context. This value profile is the primary object that users explore in the dashboard and that feeds into the cross-relational context comparison.

We also tested a temporal recency decay factor in early pilots similar to those found in previous annotation systems~\cite{shaikh2025gum}. However, it did not meaningfully improve the results. Details are in  Appendix \S\ref{sec:appendix_annotation}.

\subsection{Interactive Visualization}

\label{sec:visualization}


The \vf{} dashboard provides four primary visualization views: profile views for examining a single relational context, an evidence layer that exposes the LLM's reasoning, a comparison mode that overlays profiles across relationships, and archetype cards that characterize clusters of similar relational contexts. In pilot testing with five users, we observed that participants moved between wanting a high-level summary, fine-grained distributional detail, temporal patterns, and access to the underlying evidence. These four distinct views address these different modes of engagement.

\vspace{2pt}
\noindent\textbf{Profile views.}
Three complementary visualizations let users examine a single relational context at different levels of granularity. A {radar chart} displays all ten values as a circular fingerprint, with axes proportional to the weighted mean and shaded bands for the 95\% confidence interval (Figure~\ref{fig:cross_comparison}a). This view reveals the overall \emph{shape} of a relational context: whether someone presents as broadly balanced across values or sharply skewed toward a few. {Violin plots} show how value ratings vary across contexts (Figure~\ref{fig:cross_comparison}b). Where the radar chart shows an aggregate, the violins reveal \emph{consistency}: a tight distribution indicates stable value expression across conversations, while a wide spread suggests fluctuation. A {time series} view plots value ratings of each conversation unit within a context chronologically with LOWESS smoothing and confidence bands (Figure~\ref{fig:cross_comparison}c). This view shows whether certain values strengthen or attenuate as a relationship evolves. Interactive tooltips throughout display Schwartz's value definitions to aid interpretation.


\vspace{2pt}
\noindent\textbf{Evidence layer.}
Aggregate profiles risk appearing as black-box outputs. During pilot testing, participants frequently asked \emph{why} a particular value was rated as it was---they wanted to verify the system's inferences against their own memory of specific exchanges~\cite{liao2020questioning}. Our system provides a conversation list, filterable by insightfulness and value salience, that connects each profile to the  conversation units that produced it. Selecting a unit displays the full transcript alongside its per-value ratings and the LLM's one-sentence justifications (Figure~\ref{fig:cross_comparison}).

\noindent\textbf{Cross-context comparison.}
The comparison mode (Figure~\ref{fig:cross_comparison}) lets users overlay value profiles from different relational contexts. \vf{} produces overlaid radar charts, side-by-side violin plots ranked by differences within contexts, and time-series. Together, these place a person's different relational ``faces'' in a single display where those are comparable.

\begin{figure}[h]
  \centering
  \includegraphics[width=0.48\textwidth]{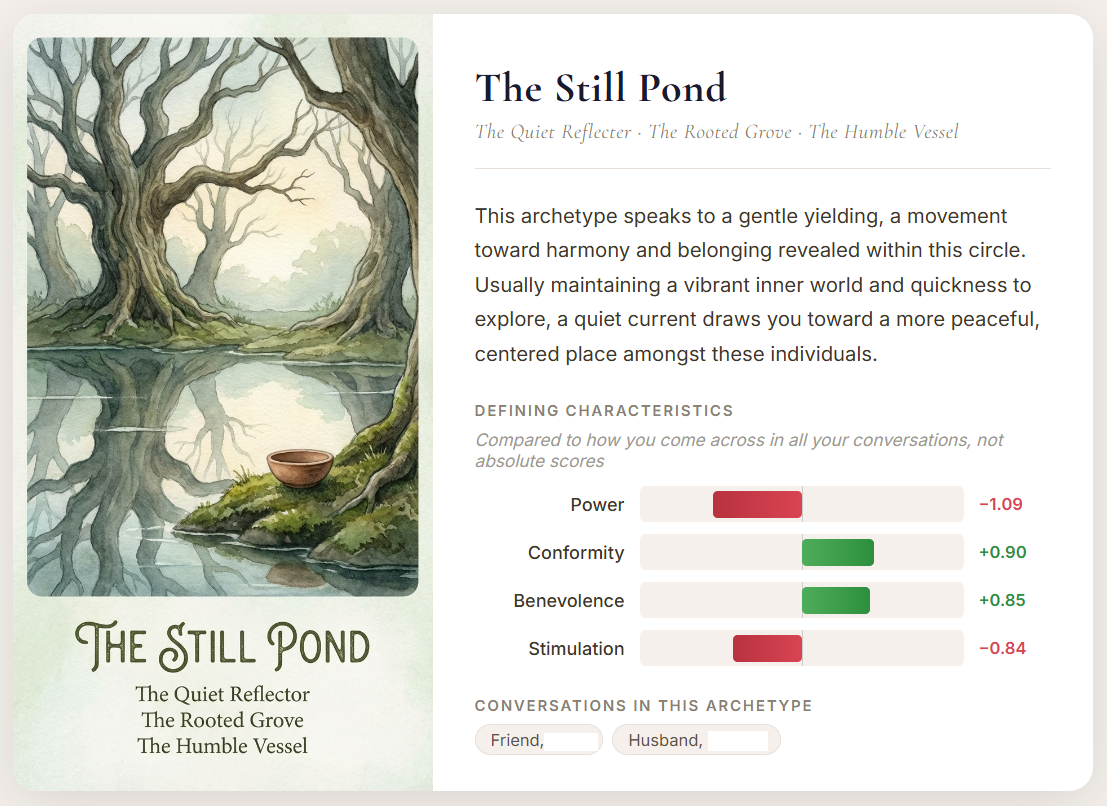}
  \caption{%
    \vf{} identifies two relational contexts with similar value sets and clusters them. The system also generates Archetype Card embodying user's demonstrated value within the cluster.
  }
  \Description{Archetype card titled "The Still Pond". The left side shows a watercolor illustration of trees reflected in a quiet pond, with a small bowl resting on the bank. Beneath the illustration are the card title and three subtitles: The Quiet Reflector, The Rooted Grove, and The Humble Vessel. The right side has text describing harmony and belonging within this group of relationships. A signed horizontal bar chart shows this archetype's defining value differences relative to the user's overall conversational profile: Power at -1.09, Conformity at +0.90, Benevolence at +0.85, and Stimulation at -0.84. Red bars extend left for negative differences, and green bars extend right for positive differences. Two labels at the bottom identify the included relational contexts as Friend and Husband, with names redacted.}
  \label{fig:cluster}
\end{figure}

\noindent\textbf{Value-face Clustering and Archetype Cards.}
Beyond user-driven comparison, the system identifies broader groupings among a user's relational contexts. It represents each face as a 10-dimensional vector in value space and groups similar ones using hierarchical clustering~\cite{murtagh2012algorithms}. To reduce noise from sparse or missing values, we impute missing entries with per-value averages and reduce the space to two dimensions via Principal Component Analysis before clustering. The number of clusters 
is determined by the largest jump in merge distances, with the merge threshold set at its midpoint following prior work~\cite{thorndike1953family}.

For each resulting cluster, the system identifies \emph{defining values}: values where the cluster's average deviates from the user's overall baseline by more than 0.5 standard deviations ($|z| \geq 0.5$). These capture how a person's value profile within that cluster diverges from the user's general communicative value baseline (Figure \ref{fig:cluster}). 

These deviations can vary in number, direction, and magnitude across clusters, which makes them difficult to interpret as a coherent whole. To make these distributed value shifts legible, we represent each cluster as an archetype card. This format draws on oracle cards, an established genre for embodying psychological archetypes as named personas in reflective and self-knowledge contexts. This is a convention that has also been adopted in technology design, as in the Tarot of Tech~\cite{tarotcardsoftech, prock2026interpretive}. We adopt this framing because it communicates distributed value patterns as a recognizable \emph{type of person} rather than a vector of scores. 

We generate each card in \textbf{\textit{two}} steps. First, we prompt Gemma-3-27B~\cite{gemma_2025} with (a)~the user's baseline value profile (per value averages across all of their conversations, between 1 to 5), and (b)~the cluster's defining value shifts, specifically which values are relatively higher or lower than the baseline and by how much. We instruct the model to produce a JSON object containing a card name, aliases, and a short evocative description, providing it with a  one-shot example modeled on published oracle decks. Stylistic guardrails direct the model to use relative, natural language instead of referencing these numeric ratings in the output. Once we have this JSON output, we pass the metadata to \texttt{gemini-3-pro-image-preview}, which produces the card's illustration. Since this is an external, proprietary image model, we do not pass any raw message text, PII, or value profile data to this model; we only give it the parsed and processed card description generated by our locally-hosted model. 



The resulting view (Figure~\ref{fig:cluster}) presents each archetype card alongside its defining value shifts, the set of interlocutors it contains and representative conversation units that best exemplify these shifts.

\section{Evaluation Methods}
\label{sec:evaluation}

The system described in \S\ref{sec:system} produces demonstrated value profiles and an interactive dashboard for all relational contexts a participant uploads. 
We evaluate the system through two quantitative analyses (\S\ref{sec:quant-eval}) and a qualitative analysis of how participants interpret and react to their value profiles (\S\ref{sec:qual-eval}). We evaluate the system end-to-end against participants' independently reported perceptions, so the validation assesses all components jointly, including the value classification. We do not separately validate the value classification of participants' conversations with human annotators, as doing so would expose participants' private conversations to third-party readers. Annotators would also lack important context and history between the interlocutors that guide the interpretation of these value ratings (\S\ref{sec:finding:qual:situated_reading_of_values}), making their labels unreliable as ground truth. The best available judge for this task therefore is the participant themselves, which motivates the setup of our study.

\subsection{Study Design}
\label{sec:study-design}

\subsection*{\textbf{Participants.}}
We recruited \N{} participants through convenience sampling via personal networks, university mailing lists, and messaging groups. Eligibility criteria required that participants be at least 18 years old, proficient in English, regular users of text-based messaging platforms (WhatsApp, Discord, Slack, or similar), and not located in the UK or EU to satisfy data-handling restrictions under our IRB protocol. Participants received \$30 USD via gift card upon session completion. 

Of the \N{} participants, 12 identified as male and 6 identified as female. The median age was 19.5 (range: 18--51), and all of them had completed or were pursuing an undergraduate degree. For a more complete demographic breakdown, see Appendix~\ref{sec:appendix_demographics}.
Each participant uploaded between 3 and 7 chat logs (median 4), for a total of 79 chat logs across iMessage, WhatsApp, Discord, and Slack. 

\vspace{2pt}
\subsection*{\textbf{Procedure.}}
Each session lasted approximately 45--60 minutes over Zoom, with sessions recorded and transcribed using Zoom's built-in transcription. Participants first uploaded their chat logs through the \vf{} browser extension we developed, which exported and anonymized conversations from their chosen messaging platforms. The extension handled Discord and Slack exports; for other platforms (WhatsApp, iMessage, Snapchat), participants used those platforms' native export feature. The system then processed these logs through the full pipeline described in \S\ref{sec:segmentation}-\S\ref{sec:visualization} to produce a personalized dashboard for each participant. Before seeing their dashboard, participants completed Schwartz's Portrait Values Questionnaire (PVQ)~\cite{schwartz2001extending} and a 10-item perceived differences questionnaire so that the system's output would not influence their responses. They then explored the dashboard, examining the quantitative comparison views (radar charts, violin plots, and time-series overlays across contexts; Figure~\ref{fig:cross_comparison}a--c) and the archetype cards generated for their value-based clusters. Each session concluded with a semi-structured interview. The study protocol was approved by the IRB at our institution.

\vspace{2pt}
\subsection*{\textbf{Perceived differences questionnaire.}}
\label{subsec:questionnaire_structure_description}
The questionnaire consists of 10 statements, each taking the form: ``How much do you agree with the following statement: I show more \textit{[Value~A]} toward \textit{[Context ~X]} than \textit{[Context~Y]}.'', with context being an interlocutor or a group chat. Participants respond on a 5-point Likert scale (Strongly Disagree, Disagree, Neither Agree nor Disagree, Agree, Strongly Agree). 

Each questionnaire item specifies a triple (a Schwartz value and two relational contexts, X and Y). For each participant, the system enumerates all triples and computes Welch t-statistic for each triple. Among triples with a significant difference $p< 0.05$, it selects the 10 with the largest absolute standardized difference.
Although we select the largest differences \emph{within} each participant, their magnitudes vary substantially across participants. As a result, the pooled elicited comparisons span nearly the full range of signed effect sizes, from large to small $d \in [-4.43, +2.97]$. Our evaluation therefore asks whether, when the system identifies a cross-context difference, the magnitude and direction of that inferred difference correlate with participants' independently reported perceptions. The only unpopulated region is the narrow band $|d| < 0.22$, where system detects no statistically significant differences. As we discuss in \S\ref{sec:user-validation-boundaries}, validation in this range is underdetermined for any system of this kind because people are themselves imperfect judges of how they come across~\cite{gilovich1998illusion} and subtle differences are where self-report has the least resolution. Sampling randomly would have drawn small differences and concentrated the test in this unadjudicable region, where any disagreement between user and the system would leave it unclear whether the system was detecting variation below the resolution of self-report or simply fitting noise.
Within each item of the questionnaire, the ordering of the two contexts is randomized, and the 10 items are shuffled before presentation. Participants with less than 10 significant comparisons received fewer questions, for a total of 174 responses from 18 participants.


\textit{Semi-structured interview.}
Following the questionnaire and dashboard exploration, we conducted an interview study asking participants to reflect on their value profiles, the patterns they noticed, and their experience with the system.

\subsection{Quantitative Evaluation}
\label{sec:quant-eval}

The discrimination analysis \S\ref{sec:self-coherence} tests whether a participant's demonstrated value profiles differ across relational contexts. The alignment analysis \S\ref{sec:alignment} tests whether those differences align with how participants perceive their own value expression across relationships. Recall that participants report these perceptions before seeing any system output, so this analysis serves as an end-to-end validation of the pipeline, which jointly assesses conversation segmentation, annotation, and weighted aggregation into profiles. We also evaluate whether our value annotations preserve the circumplex structure of Schwartz's framework (\S\ref{sec:schwartz-alignment}).

\subsubsection{Context Discrimination}
\label{sec:self-coherence}

Given an unseen conversation unit, can the pipeline assign it to the correct context based on its value profile alone?

\textit{Classification procedure.}
Each annotated conversation unit is represented as a 10-dimensional vector with one entry per Schwartz value, as described in \S\ref{sec:aggregation}. For each participant, we partition conversation units per context into a training set (70\%) and a test set (30\%) using a fixed random seed. We stratify the test set so that each context contributes an equal number of held-out conversation units, determined by the context with the fewest segments. From the training set, we compute a \textit{prototype} for each context: the centroid of the training vectors. We classify each held-out conversation unit by assigning it to the context whose prototype is nearest using cosine distance.

We exclude ratings marked ``No data'' when accounting for contribution towards each relational context's value profile. We omit conversations with less than three ratings which often contain very little interpretable text, such as exchanges consisting entirely of images.

\textit{Model specification.}
We model each classification decision as a Bernoulli trial. Let $y_{ij} \in \{0, 1\}$ indicate whether the classifier correctly assigned the $j$-th held-out conversation unit from participant $i$ to its true relational context. Because different participants may upload different number of contexts (changing the chance baseline), we fit a generalized linear mixed model:
 
\begin{equation}
\label{eq:self-coherence-glmm}
\text{logit}\, P(y_{ij} = 1) =  \text{logit}\, (1 / k_i)+ \beta_0 + u_i, \quad u_i \sim \mathcal{N}(0, \sigma^2_u)
\end{equation}
 
\noindent where $k_i$  is the number of contexts for participant $i$, so $ 1 / k_i$ is the chance probability of correct classification by random assignment. $\text{logit}\, (1 / k_i)$ encodes this as an offset on the log-odds scale. 
$u_i$ denotes per-participant random intercept.
A positive and significant $\beta_0$ indicates classification accuracy exceeds each participant’s chance level ($1/k_i$) on average. We report $\beta_0$ with a 95\% confidence interval.

We fit the model on a single canonical seed as the primary analysis. To assess sensitivity to the train--test partition, we re-run the process across 100 random seeds and report the median and interquartile range of classification accuracy. 



\subsubsection{Alignment with User Perception}
\label{sec:alignment}

\hfill\\
\textbf{System-side measure.}
As noted in \ref{subsec:questionnaire_structure_description}, each perceived difference question specifies a triple: a Schwartz value, and two relational contexts ($X$ and $Y$). For that value, the pipeline has unit-level ratings for both context $X$ and $Y$. We compute a standardized difference between the two distributions: $d = (\bar{x}_X - \bar{x}_Y) / s_{\text{pooled}}$ where $\bar{x}_X$ and $\bar{x}_Y$ are the mean scores of that specific value for relational context $X$ and $Y$, respectively, and $s_{\text{pooled}}$ is the pooled standard deviation. The sign of $d$ encodes the predicted direction (positive when the system rates the value higher for $X$ than $Y$); the magnitude encodes how strongly the system distinguishes the two.

\textbf{User-side measure.}
The participant's 5-point Likert response provides a directional judgment. Agree or Strongly Agree endorses the direction stated in the question ($X > Y$ on that value); Disagree or Strongly Disagree endorses the opposite.

\textbf{Model specification.}
We fit a cumulative link mixed model (proportional odds). Let $y_{ij} \in \{1, \ldots, 5\}$ be the Likert response to the $j$-th perceived-difference question from participant $i$:
 
\begin{equation}
\label{eq:alignment-clmm}
\text{logit}\, P(y_{ij} \leq c) = \alpha_c - \gamma_1 d_{ij} - v_i, \quad v_i \sim \mathcal{N}(0, \sigma^2_v)
\end{equation}
 
\noindent where
\begin{itemize}[nosep]
    \item $c \in \{1, 2, 3, 4\}$ indexes the four boundaries between adjacent Likert categories, each with its own threshold $\alpha_c$;
    \item $d_{ij}$ is the signed system effect size for that triple;
    \item $v_i$ is a per-participant random intercept.
\end{itemize}
 
\noindent A positive $\gamma_1$ means that as the system infers a larger difference in the direction $X > Y$, users give higher Likert responses (toward Agree), and vice versa. Formally, a one-unit increase in $d_{ij}$ increases the log-odds of responding in a higher Likert category by $\gamma_1$. We can therefore use $\gamma_1$ to measure the extent to which our model's predicted demonstrated values align with the user's perceived demonstrated values. 

\subsubsection{Annotation Alignment With the Schwartz Framework} 
\label{sec:schwartz-alignment}
To test whether our annotations behave consistently with the structure of the Schwartz value framework, we analyze whether the annotated values expressed within conversations preserve its circular structure. We correlate each pair of demonstrated values across all annotated conversation units and regress those correlations on the pair's circular distance (1 for adjacent values through 5 for opposing values). This analysis parallels the bottom-up construction of Schwartz's circumplex, whose structure was derived from empirical patterns of co-occurrence among values. We report the results in \S\ref{sec:finding:circumplex}.


\subsection{Qualitative Analysis}
\label{sec:qual-eval}

We analyzed the semi-structured interviews using a hybrid deductive-inductive thematic analysis~\cite{fereday2006demonstrating,xu2020applying}. We used deductive coding against our questions on cross-context variation and value recognition, and inductive coding for how participants interpreted and responded to their profiles. Two researchers jointly coded the first few transcripts before coding separate sets of transcripts.Throughout the coding process, they routinely shared code examples and updated the codebook to resolve disagreements and ensure consistent application of codes to the transcripts. We then pooled the codebooks, clustered the codes into candidate themes, and refined them across team discussions, consistent with guidance in HCI~\cite{mcdonald2019reliability}.

\section{Findings}
\begin{figure*}[!t]
\begin{minipage}[t]{0.48\textwidth}
  \centering
  \includegraphics[width=1.10\linewidth]{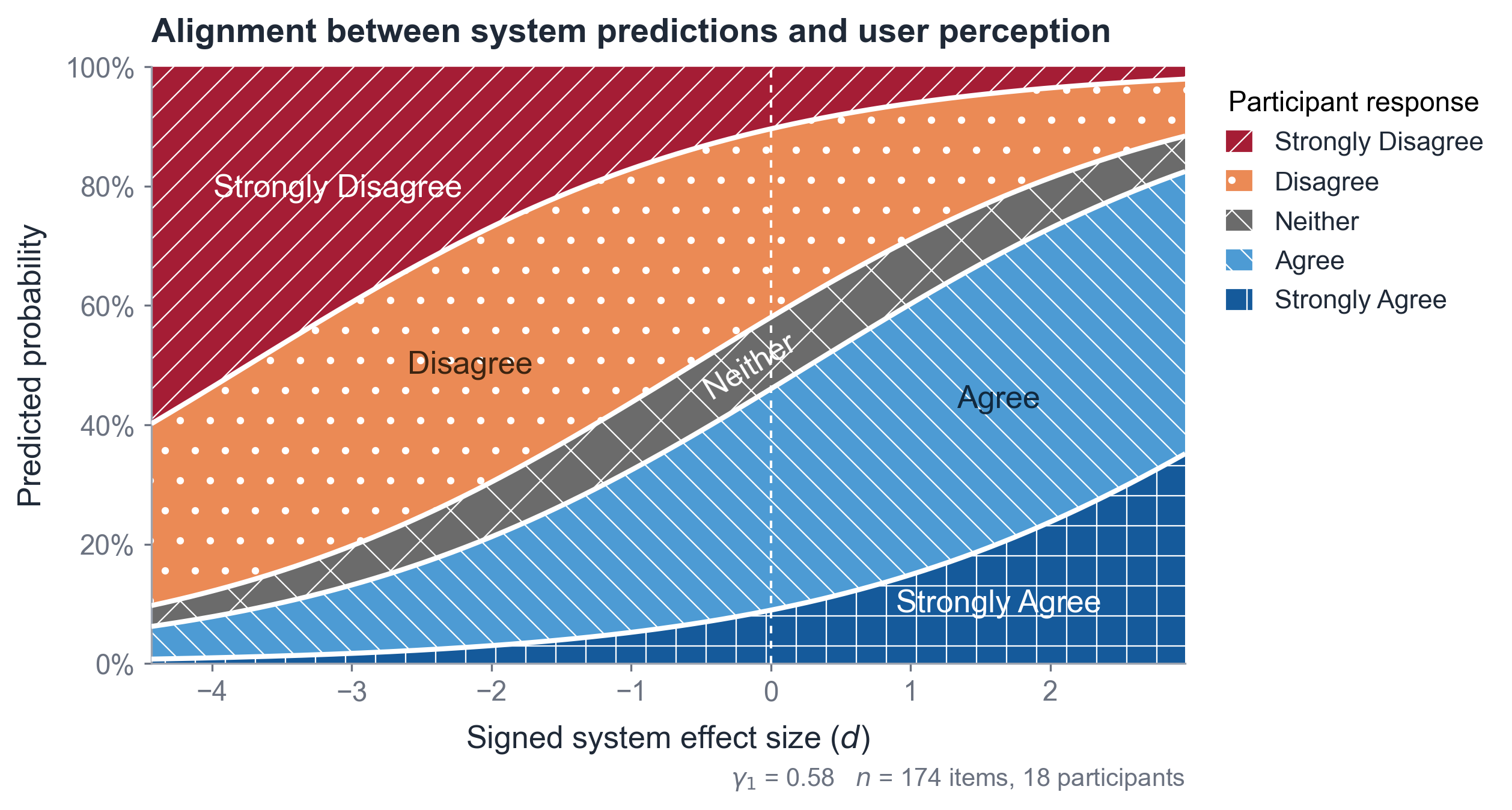}
  \caption{Alignment between system predictions and user perception. The
  stacked bands show the CLMM's predicted probability of each Likert
  category as a function of the system's effect size $d$. At negative $d$
  (system predicts $Y > X$), disagreement dominates; at positive $d$
  (system predicts $X > Y$), agreement dominates.}
  \label{fig:alignment}
  \Description{Stacked area chart shows model-predicted alignment between system differences and user perception. We plot predicted probability from 0 to 100 percent on the vertical axis against signed system effect size $d$ on the horizontal axis, with five bands representing Strongly Disagree, Disagree, Neither, Agree, and Strongly Agree. Disagreement occupies most of the chart at negative effect sizes, while agreement occupies most of the chart at the positive effect sizes. The neutral band remains very narrow throughout, but is thickest in the middle.}
\end{minipage}
\hfill
\begin{minipage}[t]{0.44\textwidth}
    \centering
    \includegraphics[
        width=\linewidth,
    ]{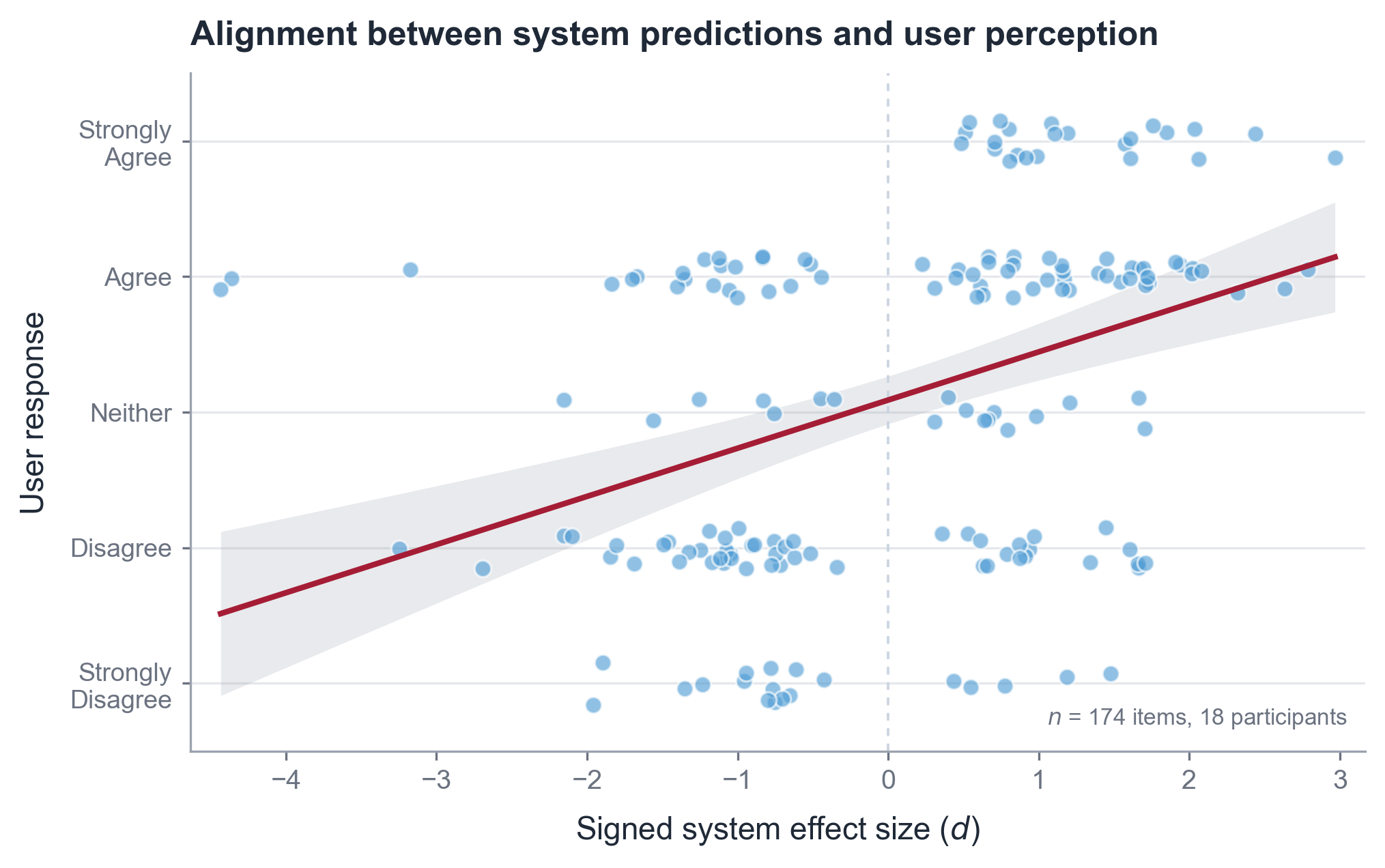}
    \caption{Observed Likert responses plotted against the system's standardized difference $d$. Each point is one perceived-difference question answered by one participant. The OLS line is fitted to the user responses coded 1 to 5 (Strongly Disagree to Strongly Agree), with a shaded region describing its conventional 95\% confidence band. (Note: actual inferential results came from the CLMM, in Figure~\ref{fig:alignment})}
    \label{fig:alignment-raw}
    \Description{Scatterplot shows observed participant responses against the system's signed effect size $d$. The five response categories on the vertical axis (from Strongly Disagree at the bottom to Strongly Agree at the top) are plotted against the signed system effect sizes on the horizontal axis. Each point represents an individual questionnaire response (jittered). A fitted line slopes upward through a 95\% confidence shaded band, showing that responses tend toward greater agreement as the system's effect size increases. Responses remain spread across multiple categories on both sides of zero.}
\end{minipage}
\end{figure*}

\subsection{Quantitative Findings}
\label{sec:finding:quant}
\subsubsection{\textbf{Value Profiles Significantly Distinguish Relational Contexts}}
\label{sec:finding:discrimination}
The classifier assigned held out conversations to the correct relational context with an accuracy of 36.8\% versus a sample-weighted chance accuracy of 23.0\% ($\beta_0 = 0.69$, OR $= 2.00$, 95\% CI $[1.62, 2.48]$, $p<0.001$). The effect holds for both genders tested separately (male: OR = 1.79, p < 0.001; female: OR = 2.64, p = 0.004). For 16 of 18 participants, the accuracy of the assignments exceeds their individual chance baseline. Across 100 random train--test splits, $\beta_0$ remains positive in every case (median 0.66, IQR $[0.61, 0.72]$). 

\subsubsection{\textbf{System-Inferred Differences Align with Participant Perceptions}}
\label{sec:finding:alignment} 
The cumulative link mixed model (Equation~\ref{eq:alignment-clmm}) yields a significant alignment ($\gamma_1 = 0.58$, OR $= 1.78$, 95\% CI $[1.43, 2.23]$, $p < .001$). This result provides end-to-end validation of \vf{} against participants' independently reported perceptions: when the system infers more strongly that a participant demonstrates a value more in one context over another, that participant is likely to independently agree. For one-unit increase in $d$, the odds of a higher response increase by a factor of 1.78. As with discrimination, the effect holds within both gender subgroups tested separately (male: OR $= 1.85$, $p < .001$; female: OR $= 1.72$, $p = .011$). 

To test whether the relationship between $d$ and response holds uniformly across category boundaries, we compared the proportional-odds model to a model with threshold-specific slopes using a likelihood ratio test. The latter did not improve model fit, supporting the proportional-odds assumption. Figure~\ref{fig:alignment} visualizes this relationship. The predicted probability of agreement (blue) grows with $d$, while disagreement (red/orange) dominates at negative $d$ values. Excluding the neutral responses, the direction that the participants selected for themselves matched the sign of $d$ in 71.2\% of the cases. Figure~\ref{fig:alignment-raw} shows the individual Likert responses plotted against $d$, displaying the variation in participants' judgments within this trend.

\subsubsection{\textbf{Demonstrated Values Preserve Schwartz's Circumplex Ordering}}
\label{sec:finding:circumplex}
Schwartz's circumplex was empirically established from how value priorities covary across people. To assess whether our demonstrated values preserve this structure, we computed pairwise correlations between all pairs of participant-normalized value ratings across all conversation units. Correlations between values declined monotonically with their angular distance from each other on the circumplex, from $r=0.26$ for adjacent to $r = -0.17$ for opposing values ($\rho= -0.54$, $p < 0.001$). As an example, conversations scoring higher on Stimulation also tended to score higher on nearby values (e.g., Hedonism and Self-Direction) and lower on an opposing value (e.g., Conformity).

Appendix Figure~\ref{fig:circumplex} visualizes the correlations across individual value pairs and the annotation distributions underlying their correlations. Appendix Table~\ref{tab:circumplex} reports the mean correlation at each circular distance.




\subsection{Qualitative Findings}

The quantitative results establish that demonstrated-value profiles vary across relational contexts, and that the direction of these differences broadly aligns with participants' own judgments of how they come across (\S\ref{sec:finding:quant}). The interviews show how participants interpret those differences once they inspect the profiles, the conversations behind them, and the visualizations.

\subsubsection{Making Sense of Variation in Demonstrated Values}

Participants recognized differences across their relationships that they had not previously articulated. P12 described writing his messages without attending to how he came across, and said that \textit{``looking at it like later, you're like, oh wow, this definitely shows like more of who I am in different circumstances.''} P9 read a low Self-Direction score in one friendship as a description of how he acts with that friend, being more supportive in that relationship. He had already thought of himself as a supportive friend, and said the profile showed him something he \textit{``knew about … but not to, I guess, this extent.''}

What participants attributed these differences to, however, varied considerably. Sometimes the shared pursuit around which a relational context was formed explained much of the profile. P2 saw this in his speedrunning Discord group where his demonstrated Achievement was high. The chat existed around speedrunning, where improving performance and completing games quickly were central to what the group was about. As he explained, \textit{``the specifics of what we were talking about.''} 
In other cases, participants tied the difference to the role they had within a relational context. For example, P7 had uploaded conversations from two group projects, leading one and working as a member of the other. The project he led showed higher Power and lower Benevolence, and the other showed the reverse. He connected this to what each project required of him, directing others and taking responsibility in one, and collaborating without that responsibility in the other.

Participants also recognized their changes in value demonstration over time reflected in the maturation of their chat logs. Within the project led by P7, Power rose across the term as he increasingly found himself guiding others. On the other hand, P16 connected a rise in Universalism in her chat with friends over the years with her shifting political position: \textit{``I went to college ... This was my transition from considering myself right-leaning to being more like I'm leftist.''}


\subsubsection{Situated Readings of Value Labels}
\label{sec:finding:qual:situated_reading_of_values}
Participants also had to determine what an abstract value label meant in the setting of a particular conversation. P18 made this distinction while comparing Security across two work relationships. A low score initially made her wonder whether something had changed in the relationship. After opening the underlying conversations, she observed that \textit{``the same value could represent a different thing depending on the context.''}. In one, it reflected the cautious language she used when asking for permission to deploy a study. In the other, it reflected her concerns about data privacy and ethical risk in the project itself.

Participants made similar translations with other values. P6 initially understood Tradition in his gaming group through the context of recurring in-game procedures and policies: \textit{``when I think tradition, we're trying to instill a rigor, instill a, group mindset''}. However, using the system to look back on his chats revealed a much more personal, casual tone that highlighted \textit{``what I think of these people versus what I'm actually chatting about in the logs''}. P5 similarly interpreted Benevolence through the norms of one close friendship, where teasing and bluntness could signal familiarity rather than lack of care. He associated a progressive increase in Benevolence with an increase in formality, saying \textit{``our friendship would be better if we were less benevolent towards each other,''} illustrating how the same value label could take on a relationship-specific behavioral meaning.

\noindent Schwartz's framework therefore supplied a common dimension, while participants supplied situated interpretation of what a value dimension referred to in a particular context.

\subsubsection{Combining Value Inference and its Evidence}
Seeing a pattern often led participants back to the conversation units behind it. When a result was surprising or unclear, participants looked to the underlying transcripts for supporting evidence or justification. Participants tended to prefer the time-series plot (Fig. \ref{fig:cross_comparison}c) particularly because each point could be interactively clicked for a trace back to the source exchange. This was particularly useful in older conversations where participants could not directly recall the subjects of discussion without the transcript. In other cases, the evidence could directly change the interpretation. P8 initially found her high Self-Direction with her boss surprising. After reading the underlying conversation units, she connected it to how much autonomy her boss gave her: \textit{``I didn't realize like how much my boss kind of lets me pick my own path.''} P6 also noted some of his value profiles had results \textit{``Completely opposite of what I answered those questions.''} After reading the annotations alongside the underlying conversations for those profiles, he reconsidered: \textit{``I can’t dispute any of the logic and the value assignments here... I wasn’t thinking about all the conversations I had with them.''}

The same evidence also gave participants grounds to disagree. P2 accepted many of the broader differences in his profiles while rejecting particular explanations (e.g., Figure~\ref{fig:cross_comparison}: d2,d3). In one project conversation, the annotation interpreted his effort to complete a task as concern for the group's welfare. He explained that he \textit{``just wanted to get things done.''} P9 meanwhile rejected a low Universalism rating in his conversations with his mother because the exchange behind it was a joke, he said, that the annotation had read at face value. In several cases, participants treated such errors as local to a conversation while still finding the aggregate differences between relational contexts plausible.

\subsubsection{Relating Demonstrated Values to Internally-held Values}
Participants also compared the relational profiles with how they understood their own values more generally. P4 said his longest relationship came closest to his PVQ profile, and attributed that to how long it had run and how much of his life it covered. P1 noticed that his value for Achievement did not seem to come across in his online chats: \textit{``I value Achievement way more than I expressed in conversations.''}

P18 treated the two as answering different questions. She read the dimensions that varied most across her relationships as describing how she treats different people, and the dimensions that stayed consistent as describing herself. She therefore wanted to compare each demonstrated-value profile with her PVQ profile, not just with other relationships.

\subsubsection{What Conversational Text Leaves Out}
\label{sec:finding:blind_spot}
For some relationships, interactions also happened outside text and therefore the uploaded chat histories captured only part of the relationship.
P2 explained that much of his Discord use was through voice chat not captured by our system. P12 texted his peers but called his father for anything substantive. P3 summarized: \textit{``It's accurate for the information it has, but it's missing information since I call people or meet in person for important conversations.''} The same relationship could also leave different traces on different platforms. P9 described Discord as the main place where he actually talked with one friend, while iMessage was used more to arrange when they would meet.

Even when the relevant exchange was present, part of its meaning stayed outside the chat logs \vf{} used. Participants pointed to images, shared history, and knowledge of events the system did not have. P10 found that apparently threatening and religious language in one exchange came from Photo Booth images and an inside joke about a film (The Exorcist). From the transcript alone, the language supported a literal reading that no one in that conversation would have made. 

\subsubsection{Supporting Self-reflection}
Participants also considered what they might do with these insights beyond the study. P18 connected lower Self-Direction to repeated apologetic wording and said the result might make her more attentive when composing future messages: \textit{``Probably next time when I send to my parents … I might even check a little bit, like, oh, did I use apologize again?''} P13 said that seeing the data laid out plainly could \textit{``confront your biases that you have about yourself, or … reaffirm them.''}

Participants also imagined this kind of reflection becoming part of settings where communication already happens. P13 wished for a plug-in embedded directly into communication platforms rather than the export-and-analysis process our study required. P6 imagined a broader organizational use, similar to existing team-development exercises, where people could examine the values and interaction patterns that characterize how they work together.
\vspace{-4pt}
\section{Discussion}

People have long turned to external frameworks to better understand and characterize aspects of themselves, from validated psychological instruments to popular personality typologies and interpretive practices such as astrology and tarot. Despite their very different epistemic foundations, these approaches share a basic form. They take some information about a person and return a representation through which that person can inspect and interpret themselves. In this work, we explore what this practice can look like when the input is instead the rich traces people accumulate through everyday digital life. LLMs make it possible to transform these traces, which were produced in the course of ordinary interaction, into structured representations of how people have presented themselves across relationships and over time. This makes an otherwise unwieldy record of everyday interaction available for self-understanding.

This shift in source material also changes what kind of self-knowledge becomes possible. A questionnaire generally asks a person to characterize herself as a whole. Conversational traces instead preserve situated expressions of the self, capturing how she responded to a friend, collaborated with colleagues, or changed within a relationship. \vf{} uses demonstrated values to make these relational patterns available for reflection without collapsing them into a single account of who the person ``really'' is.



\subsection{Values as the Analytical Substrate.}
Why use values as signals to infer, label, and compare rather than leave it to the LLM to decide what signals to label in conversations? One reason is that comparison requires a shared, quantifiable standard of measurement. An LLM doing unstructured analysis might surface patterns that are interesting, but idiosyncratic, and hard to compare and reason about systematically. Values as a substrate give a shared, quantifiable vocabulary that makes cross-context comparison possible---and one that individuals could make personally meaningful. Despite being grounded in a fixed set of dimensions, the motivational framing of value labels left enough conceptual room for our participants to interpret them within their own relational contexts (\S\ref{sec:finding:qual:situated_reading_of_values}). A second reason is that
values are designed to span a wide range of human motivational concerns and are based on decades of cross-cultural validation~\cite{schwartz1992universals, schwartz2012overview}. 
The circumplex structure of Schwartz's framework also specifies how the values relate to one another, providing a structural check on our annotations. Although our pipeline rates the ten values independently, the resulting ratings preserve the circular ordering between values that Schwartz's work established, with opposing values covarying negatively and adjacent values positively (\S\ref{sec:finding:circumplex}).

Our framework could also be adapted to other substrates, for instance emotionality or affect, or to alternative value taxonomies beyond Schwartz. We chose values in particular  because  they capture a dimension of self-presentation that is especially relevant to relational variation. Different relationships  give different priorities occasion for expression, making visible how what a person foregrounds and backgrounds shifts across relationships.
Whether other substrates produce profiles that are as personally meaningful to participants is a question for future work.

\vspace{2pt}
\subsection{Demonstrated Values and Schwartz Values.}
Demonstrated values are not intended as context-specific estimates of a person's underlying value profile. Schwartz's values are typically measured as general motivational priorities through self-report, whereas Value Faces asks what becomes expressed within particular relational contexts. Our interviews suggest that participants understood this distinction. They could regard a value as personally important while recognizing that it was not strongly expressed in a particular conversation, and some explicitly wanted to compare their Schwartz PVQ profile against their different demonstrated-value profiles.

Comparing the two can help people examine which of their broader priorities carry across relationships and which shift with the roles, concerns, or demands of a particular relationship.
Neither agreement nor divergence would establish that one profile is a more authentic account. For systems that infer identity-related constructs from behavior, observed expression should not automatically be treated as a noisy proxy for an underlying trait.

\vspace{2pt}
\subsection{Behavioral Inference for User-Defined Ends.}
Proprietary platforms have long analyzed users' behavioral traces, including messages, interactions, and engagement patterns, for platform-defined ends such as engagement maximization~\cite{zuboff_surveillance} and ad targeting~\cite{liu2021machine,ur2012smart}. 
Until recently, such higher-level inference over everyday data was mostly available through complex models built and operated by platforms.
LLMs change this landscape because they can reason over the semantic content of heterogeneous traces without requiring task-specific training data. This general reasoning capacity can be directed toward user-defined ends.

This opens new possibilities for personal informatics. Many personal informatics systems begin with an intention to observe oneself. People decide what they want to understand and collect or track the corresponding data~\cite{epstein2015lived,elsden2016quantified}. Yet people already leave extensive behavioral traces through ordinary digital activity, including what they write, watch, search for, purchase, and interact with, without generating these traces for the purpose of self-tracking. LLMs make it possible to apply interpretive frameworks retrospectively to such traces, allowing people to examine aspects of themselves they did not set out to measure when the data were created~\cite{shaikh2025gum}.

In this work, we instantiate this possibility using conversational histories. By interpreting these traces through the framework of human values, \vf{} allows people to inspect patterns in how they have presented themselves across relationships and over time. Demonstrated values therefore illustrate how behavioral inference can be repurposed from learning about users for platform-defined ends toward helping people learn from their own behavioral histories.
One direction for future work is to explore interfaces that facilitate acting on demonstrated values as well as reflecting on the effects of doing so.

\vspace{2pt}
\subsection{Making Inferred Profiles Legible.}
The dashboard presents inferred value profiles through both quantitative views and an archetype representation. The visualizations let participants compare relational contexts, inspect variation, and trace change over time. These views assume some comfort with distributional reasoning, but they also gave participants direct access to the structure of the inferred value profiles. Participants used the higher-level views to notice differences across relationships and over time, and also inspected the annotation when a pattern was surprising or unclear. This sometimes made an inference more understandable, and in other cases exposed missing context or an annotation error. 

Through archetype cards, we wanted to provide a more cognitively accessible representation of relational contexts with similar demonstrated-value profiles (\S\ref{sec:visualization}). 
In our study, however, participants engaged more with the value scores and the evidence layer than with the aggregated archetypes. This may reflect that our participants largely had technical backgrounds and were comfortable reasoning from quantitative data. The archetypes could be more valuable for populations less comfortable with data-driven interfaces, which we leave as a question for future work.

\vspace{2pt}
\subsection{Defining Relational Context.}
\vf{} treats each interlocutor or group as a relational context, but that context encompasses more than audience identity alone.
People do not choose platforms or take on roles independently of whom they are interacting with. These choices, in turn, shape what gets discussed and expressed.
This is consistent with Goffman's account of audience, role, and setting as jointly constituting an interactional whole~\cite{goffman1959presentation}. In digital communication, platform becomes part of that setting. Its norms and affordances shape what kinds of interaction are easy or expected, while people may also use different platforms for different relationships or for different kinds of interaction within the same relationship. P9, for example, used Discord to talk with one friend but iMessage primarily to arrange when they would meet. Role can similarly shape what arises in an interaction. P7's two group projects both involved peers of similar backgrounds, but he occupied different roles and attributed the different value patterns partly to those roles. Differences between demonstrated-value profiles should therefore be interpreted in relation to the audience, role, platform, and purpose of each interaction.


\vspace{2pt}
\subsection{The Boundaries of User Validation.}
\label{sec:user-validation-boundaries}
We validate our system's inferences against user perception across the range of effect sizes represented in our evaluation. Our evaluation does not cover the narrow region where system-inferred differences are not statistically significant ($|d| < 0.22$). 
In this regime, it is unclear whether user perception can serve as ground truth. If the system infers a subtle difference that a user does not perceive, two interpretations are possible.
Either the system is detecting genuine variation that falls below the threshold of conscious self-awareness, or it is overfitting to noise in the conversational data. The former is plausible since people are not always reliable reporters of subtle differences in how they present themselves~\cite{carlson2009evidence}, which is part of why inference over their conversations is useful in the first place. But this explanation also means that in the low-effect-size regime, user perception can no longer serve as a ground truth. The validity of the system's inferences in this regime therefore remains an open question.

\section{Limitations and Future Work}
\noindent\textbf{Participant-curated Data.} Participants chose which conversations to share with us, and several acknowledged during pilot sessions that they held back conversations they considered too sensitive, even after learning that raw text would not be retained. Our user study may have operated on a curated subset of each person's relational landscape, and the profiles may underrepresent relationships where value expression is strongest. Our privacy architecture (\S\ref{sec:privacy}) reduces some of these concerns but does not eliminate them. Some participants also needed guidance during export, upload, and dashboard navigation. A native platform plugin could reduce friction and simplify these steps.

\vspace{2pt}
\noindent\textbf{Text-only Communication.}
\vf{} captures only text-based messaging. Although we require at least two months of history and ten conversation units for each uploaded chat log, participants noted that their most substantive interaction in some relationships took place over calls or in person. In those cases, the demonstrated-value profiles represent only the part of the relationship expressed through messaging.

\vspace{2pt}
\noindent\textbf{Pipeline Validation.} 
We designed the individual components of the pipeline, including conversation segmentation, insightfulness scoring, evidence weighting, and LLM value annotation, using prior work~\cite{kolluri2025alexandria, epstein2025measuring, jahanbakhsh2025value} and iterative piloting but did not independently validate each component ourselves. Our end-to-end analysis found that participants' independent assessments of their own value expression across relationships tended to align with \vf{}'s inference. The contribution of each component to these results, and whether alternatives would improve alignment, remains a question for future work.

We used a two-hour time-gap heuristic (\S\ref{sec:segmentation}) to segment asynchronous messages and found that the value profiles were robust to alternative time-gaps (\S\ref{sec:appendix_segmentation}). Future work could explore more sophisticated dialogue segmentation methods and evaluate whether they produce units that better reflect conversational structure.

We did not use expert human annotators in our evaluation (\S\ref{sec:quant-eval}) because doing so would expose participants' private conversations to third-party readers, with privacy risks that de-identification might not adequately address. External annotators would also lack the relational history participants drew on when interpreting value ratings (\S\ref{sec:finding:qual:situated_reading_of_values}). The best available judge for this task, then, is the participant themselves, which motivates their uses in Sections \ref{sec:quant-eval} and \ref{sec:qual-eval}.

\vspace{2pt}
\noindent\textbf{Sample.} 
Our participant pool primarily consisted of young, university-educated adults who were comfortable with technology and accustomed to text-based messaging. The sample is modest in size and skewed toward male participants (66\%). We found statistical significance within each gender subgroup (\S\ref{sec:finding:discrimination}). However, it remains an open question as to whether the results are generalizable to other user populations with different messaging norms.

\vspace{2pt}
\noindent\textbf{Annotation quality.} 
Our local 27B model occasionally missed contextual nuances or misinterpreted sarcasm as literal value expressions (\S\ref{sec:finding:blind_spot}). Sarcasm generally yields low inter-rater consistency among LLMs~\cite{bojic2025comparing}, and such misrepresentation is particularly prevalent in smaller models~\cite{ajayi2026humorrank}. Our evaluation (Appendix \S\ref{sec:appendix_model}) suggests that larger models could handle these nuances better. A future version could additionally allow users to correct annotations that misrepresent their conversations.

\noindent\textbf{Subjective Value Interpretation.}
Our system applies a common interpretive lens to values manifested in text. This yields a single reading that may diverge from the user's or their interlocutor's perspective~\cite{epstein2025measuring}. For applications centered on self-reflection, presenting an alternative reading rather than mirroring the user's interpretation can be advantageous, as it prompts the user to consider, contest, or incorporate external viewpoints into their own understanding. Ultimately, which perspective is relevant to surface depends on the purpose of the system, an area we leave open for future work.

\section{Conclusion}
This paper introduced \emph{demonstrated values}: the values a person expresses within a specific relational context, serving as a behaviorally grounded complement to self-reported value profiles. \vf{} operationalizes this concept through a privacy-first pipeline that turns everyday chat logs into context-sensitive, interactive value profiles grounded in Schwartz's ten-value system. Our study with \N{} participants shows that these profiles are statistically distinguishable between relational contexts and are meaningful to participants, who often discovered patterns they had felt but never articulated. More broadly, our system illustrates that the behavioral traces which accumulate on digital messaging platforms contain rich signals about identity and self-presentation, and can be effectively funneled towards tools that serve user self-understanding.


\bibliographystyle{ACM-Reference-Format}


\appendix
\appendix
\section{Circumplex Ordering in Demonstrated Values}
\label{sec:appendix_circumplex}

\begin{figure*}[!t]
  \centering
  \includegraphics[width=\linewidth]{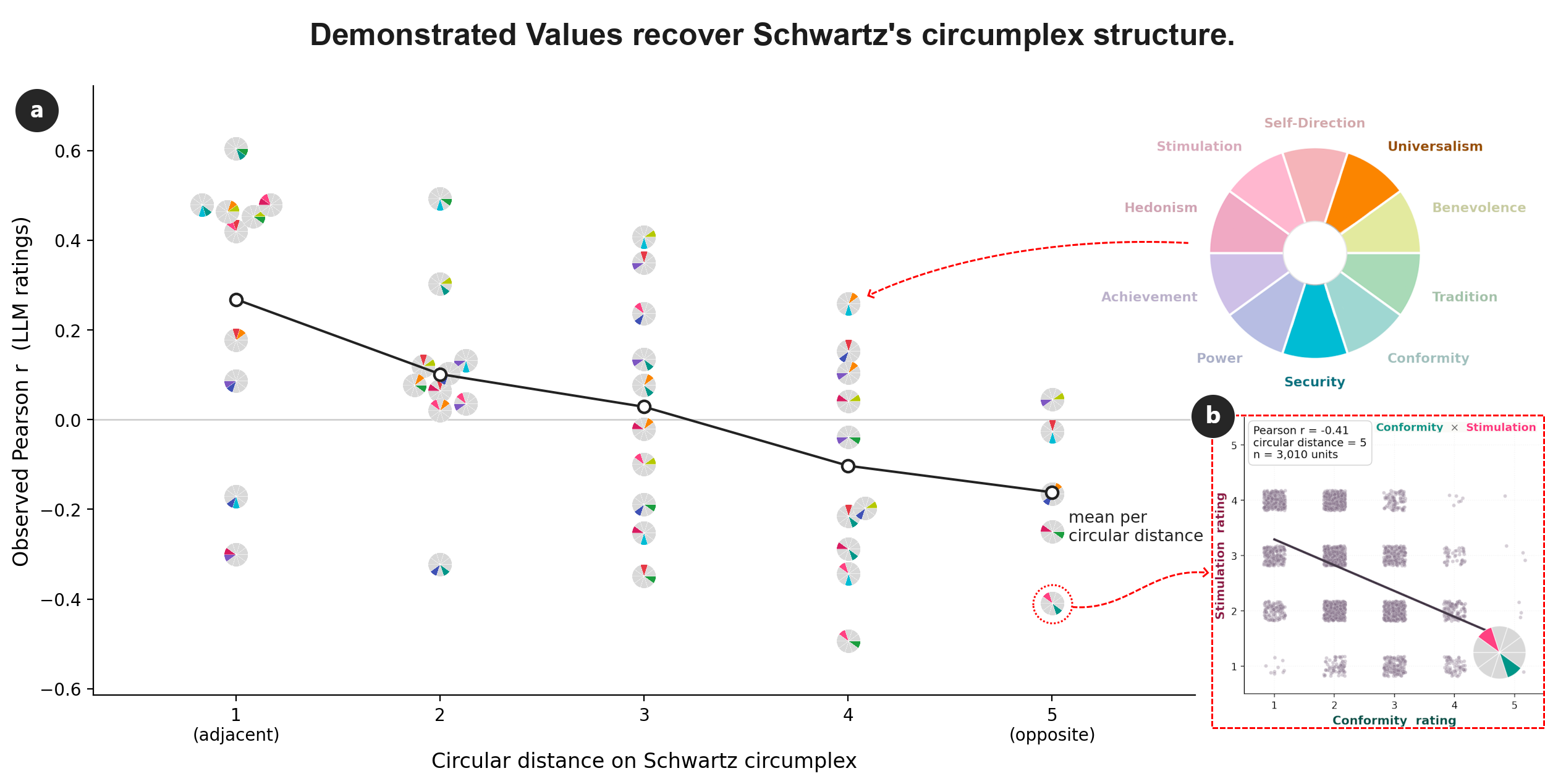}
  \caption{
    \textbf{(a)} Each point represents the relationship between two of Schwartz's values, with proximity on the chart plotted on the x-axis and the Pearson correlation across annotated conversations plotted on the y-axis. Points are represented using miniature value circle whose wedges describe the specific values they represent. Values correlate most highly when adjacent and correlate most negatively when at opposite ends of the wheel.
    \textbf{(b)} For each plotted point, we calculate that value pair's correlation by collecting all value annotations from our study which contain both of those values and statistically testing for association. Because all annotations are rated on integers 1--5, we jitter the annotation points for visual clarity, and draw the line of linear fit.
  }
  \Description{Two-panel figure which shows correlations between demonstrated values in relation to their position on Schwartz's circumplex. (a) plots value-pair correlation on the vertical axis, and circular distance from 1 (adjacent) to 5 (opposite) on the horizontal axis. (b) expands an opposing pair of values into a scatterplot of conversation-level ratings (jittered), Conformity on the horizontal axis and Stimulation on the vertical axis, with a fitted line sloped downward. This value pair has a Pearson correlation coefficient of negative 0.41 across 3,010 conversational units.}
  \label{fig:circumplex}
\end{figure*}

\autoref{tab:circumplex} reports the mean correlation between value ratings at each circular distance, summarizing the analysis in \S\ref{sec:finding:circumplex}. The decline is monotonic in both the pooled and the within-participant centered version. Distance 5 has five pairs rather than ten because opposing pairs on a ten-point circle are counted once.

\begin{table}[h]
    \centering
    \small
    \caption{Mean Pearson correlation between pairs of value ratings, by circular distance on Schwartz's model (1 = adjacent, 5 = opposing). Computed over 5{,}869 annotated conversation units from 79 chat logs. The within-participant column centers each participant's ratings before correlating, removing between-person differences in overall rating level.}
    \Description{}
    \label{tab:circumplex}
    \begin{tabular}{crrr}
        \toprule
        \shortstack[c]{\textbf{Circular}\\\textbf{distance}} & \textbf{Pairs} & \shortstack[r]{\textbf{Pooled}\\\textbf{Correlation}} & \shortstack[r]{\textbf{Within-participant}\\\textbf{Correlation}} \\
        \midrule
        1 (adjacent) & 10 & $+0.284$ & $+0.263$ \\
        2            & 10 & $+0.104$ & $+0.091$ \\
        3            & 10 & $+0.024$ & $+0.030$ \\
        4            & 10 & $-0.126$ & $-0.099$ \\
        5 (opposing) &  5 & $-0.189$ & $-0.168$ \\
        \midrule
        \multicolumn{2}{l}{Spearman $\rho$ vs.\ distance} & $-0.553$ & $-0.575$ \\
        \multicolumn{2}{l}{$p$} & $<.001$ & $<.001$ \\
        \bottomrule
    \end{tabular}
\end{table}

The strongest adjacent correlations are Tradition--Conformity ($r = 0.61$), Self-Direction--Stimulation ($r = 0.50$), Conformity--Security ($r = 0.48$), and Hedonism--Stimulation ($r = 0.45$). Coverage varies substantially across values: Self-Direction, Stimulation, and Conformity carry a rating in over 94\% of conversation units, while Tradition and Universalism are rated in only 18\% and 21\%, so correlations involving those two rest on far fewer units.

\section{Conversation Segmentation Robustness}
\label{sec:appendix_segmentation}

To test how sensitive value profiles are to the two-hour gap threshold used for conversation segmentation, we re-segmented three chat logs at eight alternative thresholds from 0.5 to 12 hours. We selected chat logs from a one-on-one DM, a large group chat, and a Slack workspace to capture varying conversational norms and response tempos. 
These logs came from one of the authors' own messaging history.

Each new segmentation produces conversations that differ from the two-hour gap baseline: smaller gaps split baseline conversations, and larger gaps merge them. We re-annotated all conversations from each chat log and for each gap size using the same model (Gemma~3-27B-IT), prompt, and parameters as the baseline---the only difference being the segmentation threshold itself. Of 5{,}633 total conversation-threshold combinations, 4{,}401 were identical to the two hour-gap baseline (they included identical conversation turns). We then re-aggregated the 10-value profiles for each chat log and threshold (\S\ref{sec:aggregation}) and compared each threshold's resulting profile against the two-hour baseline.

For each threshold, we computed Pearson $r$ and Spearman $\rho$ between the resulting aggregate value vector and that of the two-hour baseline. \autoref{tab:segmentation} reports the results. Profiles were highly stable: mean Pearson $r$ remained above 0.99 at almost every threshold (the outlier being at 0.5 hours, with $r = 0.989$), and mean absolute deviation on the 1--5 rating scale stayed below 0.08. The closest thresholds (1.5 and 3 hours) deviated by 0.032 and 0.023 on average, respectively. Even the 12-hour threshold---which reduced the average conversation count by 30.7\%---produced $r = 0.996$.

\begin{table}[h]
    \centering
    \caption{Robustness of aggregate value profiles to conversation segmentation threshold. For each gap threshold, chat logs are re-segmented and re-annotated, and the resulting aggregate value profile is compared to the profile produced at the 2-hour baseline. Each row reports the mean across three chat logs. Pearson $r$ and Spearman $\rho$ measure correlation between re-annotated and baseline profiles; MAD reports mean absolute deviation on the 1--5 scale.}
    \label{tab:segmentation}
    \begin{tabular}{rrrrr}
        \toprule
        \textbf{Gap (h)} & \shortstack[re]{\textbf{Mean Annotated}\\\textbf{Units}} & \textbf{Pearson $r$} & \textbf{Spearman $\rho$} & \textbf{MAD} \\
        \midrule
            0.5  & 298.0 & 0.989 & 0.963 & 0.077 \\
            1.0  & 273.3 & 0.996 & 0.980 & 0.065 \\
            1.5  & 260.3 & 0.999 & 0.988 & 0.032 \\
            2.0  & 249.7 & \multicolumn{3}{c}{(baseline)} \\
            3.0  & 234.7 & 0.999 & 0.996 & 0.023 \\
            4.0  & 220.7 & 0.999 & 1.000 & 0.035 \\
            6.0  & 207.7 & 0.999 & 1.000 & 0.032 \\
            8.0  & 200.3 & 0.999 & 0.996 & 0.032 \\
            12.0 & 173.0 & 0.996 & 0.988 & 0.059 \\
        \bottomrule
    \end{tabular}
\end{table}
We find that the two-hour threshold is not a fragile parameter choice: any threshold between 1 and 8 hours produces near-identical aggregate profiles. Additionally, because the evidence-weighted aggregation averages over many conversations, it absorbs the variation introduced by merging or splitting individual segments at the boundaries.

 
\section{Annotation Prompt}
\label{sec:appendix_prompt}

\autoref{fig:annotation_prompt} shows the complete system prompt given to the LLM for each conversation.
At runtime, \texttt{\{value\_prompt\}} is replaced with the ten Schwartz value definitions listed in \autoref{tab:value_defs}, and \texttt{\{value keys\_json\}} is populated with the corresponding JSON keys.
Author names in the conversation transcript are replaced with \textsc{Target Author} and \textsc{Conversation Partner~$k$} before appending to the prompt.

\begin{figure*}[]
    \centering
    \begin{tcolorbox}[colback=gray!5, colframe=gray!50, fontupper=\small\ttfamily, title={\small\sffamily\bfseries Annotation System Prompt}, left=4pt, right=4pt, top=4pt, bottom=4pt]
        You are a psychology assistant specialized in Schwartz's Theory.
        You will be given a chat excerpt of the TARGET AUTHOR speaking within the context of one or more other conversational partners.
        You will provide a short one-sentence analysis of the TARGET AUTHOR on each of the Schwartz values, then you'll rate them on each value using a scale of 1 (doesn't demonstrate value) to 5 (demonstrates values highly) based on your best estimation of their expressed value system through their behavior within the provided conversation. For each Schwartz value, when there is no evidence for or against the value within the provided conversation, rate it a -1 (No data).\\[4pt]
        IMPORTANT: -1 (No data) and 1 (doesn't value) are NOT the same. Use -1 when the value simply was not discussed or relevant to the conversation. The absence of evidence is NOT evidence of non-value. Use 1 only when the conversation contains concrete evidence that the author actively rejects, dismisses, or acts against that value. For example, if Tradition was never mentioned, rate it -1. If the author explicitly mocked traditional customs, rate it 1. Note that the uploaded chat excerpt may contain links, videos, or images which you cannot parse and should immediately ignore as null data.\\[4pt]
        For each value, also provide an `evidence\_strength' score from 0 to 100 indicating how much textual evidence in this conversation supports your rating. 0 means you found no relevant text (rating should be -1). 100 means you found extensive, direct textual evidence (e.g., explicit statements, repeated behavioral patterns). 50 means moderate evidence, e.g.\ a few relevant remarks but nothing definitive. For values with very low evidence (the value simply wasn't discussed / didn't come up), rate the value -1 with an evidence score of 0. A low rating with high evidence\_strength means the conversation contains concrete text where the author rejects or acts against that value. A high rating with high evidence\_strength means the conversation contains concrete text where the author is embodying or expressing that value.\\[4pt]
        Afterward, you will also give an ``insightfulness'' rating on a scale of 0 (irrelevant) to 100 (introspective) that estimates how useful the conversation is for understanding the author's demonstrated system of values.
        Also provide `insightfulness\_reasoning': a one-sentence explanation of why you gave the insightfulness score you did.\\[4pt]
        \textit{\{value\_prompt\}}\\[4pt]
        You must respond with a valid JSON object matching this exact structure:\\
        \{``schwartz\_values'': \{\textit{\{value\_keys\_json\}}\}, ``insightfulness\_reasoning'': ``...'', ``insightfulness'': 0--100\}
    \end{tcolorbox}
    \caption{Complete annotation system prompt. The \texttt{\{value\_prompt\}} placeholder is replaced with the value definitions in \autoref{tab:value_defs}. The \texttt{\{value\_keys\_json\}} placeholder is replaced with the expected JSON response structure containing per-value fields for \texttt{analysis}, \texttt{rating}, and \texttt{evidence\_strength}.}
    \Description{Box titled "Annotation System Prompt" containing the instructions provided to the annotation model used within this system.}
    \label{fig:annotation_prompt}
\end{figure*}

\begin{table}[!t]
    \centering
    \caption{Schwartz value definitions provided to the LLM via the \texttt{\{value\_prompt\}} placeholder.}
    \label{tab:value_defs}
    \begin{tabular}{p{2.2cm}p{5.3cm}}
        \toprule
        \textbf{Value} & \textbf{Definition} \\
        \midrule
        Self-Direction & Independence of thought and action; choosing, creating, exploring. \\
        Stimulation & Excitement, novelty, and challenge in life; avoiding boredom. \\
        Hedonism & Pleasure and sensuous gratification for oneself; enjoying life. \\
        Achievement & Personal success through demonstrating competence according to social standards. \\
        Power & Social status and prestige, control or dominance over people and resources. \\
        Security & Safety, harmony, and stability of society, relationships, and self. \\
        Conformity & Restraint of actions, inclinations, and impulses likely to upset or harm others and violate social expectations or norms. \\
        Tradition & Respect, commitment, and acceptance of the customs and ideas that traditional culture or religion provide. \\
        Benevolence & Preservation and enhancement of the welfare of people with whom one is in frequent personal contact (the ``in-group''). \\
        Universalism & Understanding, appreciation, tolerance, and protection for the welfare of all people and for nature. \\
        \bottomrule
    \end{tabular}
\end{table}

\section{Model Selection}
\label{sec:appendix_model}

\vf{} processes private conversations, so the annotation model must be locally-hosted (see \S\ref{sec:system}). This constrains us to open-weight models, which for practical reasons must be small enough to run on a single GPU. This limits us to relatively low-performance language models. Therefore, before committing to a model, we needed to ensure our entire pipeline was robust enough to generate reliable results with different annotators. We annotated {four of our own chat logs} from four relatively-similar contexts (different friend groups) using 12 LLMs (three proprietary, nine open-weight candidates) to gauge stability of the value annotations and to evaluate the performance of candidate models. We included the proprietary models as an upper-bound reference. Since stronger models can generally be expected to perform better at annotation tasks, similarity to their outputs provides a useful proxy for annotation quality of the open weight candidates. Because this phase required accessing proprietary frontier model APIs, \textbf{we did not use participant data}. Instead, we used the chat logs mentioned above. We also explicitly obtained consent from all interlocutors within all the chats used for this phase.

\subsection{Cross-Model Agreement}

Because \vf{} foregrounds the \emph{shape} of value profiles rather than the absolute rating magnitudes, our primary question is whether models agree on the relative ordering of a person's values within each context. We assess this at two levels: raw conversation-level ratings and aggregate profiles.

At the conversation level, we compute the mean absolute error (MAE) between all pairwise model ratings on the 1--5 scale, excluding $-1$ (``no data'') annotations from both sides of each comparison. Across 12 models, 245 conversations, and 10 values (53{,}389 pairwise comparisons), we observe a pooled MAE of 0.84 ($SD = 0.79$). Models typically agree within one rating point on individual conversation annotations.

At the aggregate level, we compute each model's mean 10-value profile per context and measure pairwise Spearman $\rho$ across all model pairs (\autoref{tab:profile_sim}). Across all $\binom{12}{2} \times 4 = 264$ comparisons, we observe a mean $\rho = 0.81$ ($SD = 0.15$), indicating that models strongly agree on the shape of the aggregate value profile.

\begin{table}[h]
    \centering
    \caption{Pairwise Spearman $\rho$ between aggregate 10-value profiles across all 66 model pairs, per chat log context. For each model, the aggregate profile is the mean rating per Schwartz value (excluding $-1$ annotations) across all conversations in that context. $\rho$ measures whether models rank the ten values in the same relative order for a given person and context.}
    \label{tab:profile_sim}
    \begin{tabular}{lrrr}
        \toprule
        \textbf{Context \#} & \textbf{Mean $\rho$} & \textbf{SD} & \textbf{\# Convs} \\
        \midrule
        1 & 0.863 & 0.064 & 44 \\
        2 & 0.835 & 0.083 & 95 \\
        3 & 0.832 & 0.129 & 82 \\
        4 & 0.693 & 0.205 & 24 \\
        \midrule
        \textbf{Avg.} & \textbf{0.806} & \textbf{0.147} & \textbf{61.25} \\
    \bottomrule
    \end{tabular}
\end{table}

\subsection{Model Evaluation}
\label{sec:appendix_model_selection}

\begin{table*}[t]
\centering
\caption{Per-model annotation characteristics across 245 conversations. Models above the rule are proprietary API models included for concordance validation; models below are open-weight candidates for locally-hosting on a single GPU. \textit{Fail rate}: unparseable JSON responses. \textit{Non-trivial rate}: percentage of value ratings that are not $-1$ (``no data''). \textit{Mean var.}: within-context per-conversation variance (lower = more stable). \textit{Consistency}: cosine similarity of aggregate profiles across repeated runs with different seeds (higher = more reproducible). \textit{vs.\ GPT-5.2}: mean per-conversation cosine similarity of 10-value rating vectors against GPT-5.2's ratings of the same conversations ($-1$ values excluded pairwise).}
\label{tab:model_performance}
\begin{tabular}{lrrrrrr}
\toprule
\textbf{Model} & \textbf{Fail \%} & \textbf{Non-trivial \%} & \textbf{Mean var.} & \textbf{Consistency} & \textbf{vs.\ GPT-5.2} & \textbf{Mean insight} \\
\midrule
GPT-5.2               & 2.6  & 67.1 & 0.589 & 0.992 & ---   & 34.7 \\
GPT-5-mini            & 2.5  & 45.0 & 0.772 & 0.951 & 0.979 & 29.6 \\
GPT-4.1               & 4.8  & 53.3 & 0.700 & 0.990 & 0.986 & 38.7 \\
\midrule
Gemma 3-27B-IT        & 0.8  & 75.2 & 0.633 & 0.990 & 0.969 & 47.9 \\
Qwen 2.5-7B           & 0.8  & 52.9 & 0.799 & 0.989 & 0.934 & 46.7 \\
Mistral Small 3.1-24B & 2.0  & 60.4 & 0.863 & 0.992 & 0.973 & 33.3 \\
MiMo-v2-flash         & 4.1  & 54.2 & 0.973 & 0.952 & 0.969 & 28.3 \\
Llama 3.3-70B         & 8.2  & 50.9 & 0.757 & 0.963 & 0.972 & 39.8 \\
Llama 3.1-8B          & 9.4  & 62.6 & 1.565 & 0.989 & 0.919 & 50.1 \\
GPT-oss-20B           & 33.9 & 29.0 & 0.721 & 0.960 & 0.965 & 17.3 \\
Qwen3-32B             & 34.7 & 37.2 & 0.776 & 0.990 & 0.966 & 34.1 \\
Qwen3-14B             & 49.4 & 25.5 & 0.698 & 0.902 & 0.961 & 18.1 \\
\bottomrule
\end{tabular}
\end{table*}

Given that cross-context annotation patterns hold across model families and sizes, we selected the production model on operational criteria (\autoref{tab:model_performance}). Because the stronger proprietary models can be expected to perform better at most tasks than our open-weight candidate models, we can effectively use the annotation similarity to the proprietary models as a guideline for annotation performance.

Among the locally-hostable candidates, Gemma~3-27B-IT stood out on every operational metric. It produced the fewest unparseable JSON responses (fail rate 0.8\% vs.\ 2--49\% for other candidates), meaning the pipeline rarely had to discard its annotations. Its per-conversation unit value annotations within each chat log were among the most consistent (second only to GPT-5.2), and its annotations were stable across repeated runs with different random seeds (Consistency 0.990). Its average per-conversation agreement with GPT-5.2 (0.969) was one of the highest among open-weight models, indicating that its annotations closely tracked those of the largest proprietary model tested. 

We, therefore, selected Gemma~3-27B-IT (INT4 quantized) as the production model, deployed on a single GPU via vLLM with temperature~0 and a fixed random seed for reproducibility.

\section{Annotation Pipeline Refinement}
\label{sec:appendix_annotation}

During development, we conducted a series of tests on our own internal chat logs (4 contexts, 245 conversations) to refine the annotation prompt and aggregation strategy specifically tailored to our selected base model, as described below.

\subsection{The $-1$ vs.\ 1 Disambiguation}

Early testing revealed that Gemma~3-27B-IT frequently conflates $-1$ (``value not demonstrated'') with 1 (``demonstrates value antithesis''). It would assign a rating of 1 to 38\% of its non-null annotations, far more than would be expected of a signal to the \textit{opposition} of a value. By comparison, GPT-5.2 assigned a rating of 1 to only 3.4\% of its non-null annotations in the same conversations. Cross-referencing confirmed that 68--70\% of Gemma's rating-1 judgments were cases where GPT-5.2 reported no data at all.

We added explicit disambiguation language to the prompt: ``absence of evidence is not evidence of absence \ldots use 1 only when the conversation contains concrete evidence that the author actively rejects or acts against that value'' (full prompt in \autoref{sec:appendix_prompt}). On re-annotation with the fixed prompt, Gemma's rating-1 count dropped from 767 to 359 across the same 245 conversations (a 53\% reduction), bringing its rating-1 rate to 20\% of non-null annotations, improving alignment with user-expectations.

\subsection{Evidence Strength Scoring}
\label{sec:appendix_evidence}

Even after disambiguation, Gemma had a much higher frequency of rating 1s than stronger models (e.g. GPT-5.2), however quite a few of these annotations were qualitatively justifiable. In cases where there was only limited textual evidence for a value (e.g. deciding to procrastinate on work might demonstrate low value for Tradition/Conformity), models like GPT-5.2 would consistently rate -1, while Gemma~3-27B-IT consistently rated 1. Rather than one model being more correct than the other, these models simply had a different standard of evidence with which to assign ratings. Even among human annotators, value annotations are subjective and reflect annotators' own value orientations~\cite{epstein2025measuring}.

To handle these cases pragmatically, we asked the LLM to assign per-value confidence scores based on how much evidence for each value was present in the text. We asked the model specifically for an `evidence\_strength` score, using the framing of concrete textual evidence to encourage down-weighting partially-justifiable value ratings, and to up-weight concrete ones. Beyond serving as an aggregation weight (\S\ref{sec:aggregation}), this evidence score also greatly improved user-agreement with LLM conversation-level annotations in our pilot studies.

\subsubsection{Recency Decay Ablation}

We also tested whether a temporal recency factor (a common staple in LLM annotation systems) could further improve aggregate value profile quality. We used positional decay: a linear weight from a configurable floor (applied to the earliest conversation) to 1.0 (the most recent), which adapts to relationship duration without requiring a fixed half-life in calendar time. We swept 11 floor values from 0.0 (earliest conversation gets zero weight) to 1.0 (no decay).

The results were inconclusive. While this did effectively reduce the variation in value profiles for chat logs with a large change over time, the resulting aggregate profiles did not necessarily align better with user expectations. Our results remained nearly identical across different decay factors, so we omitted recency decay from the final system.


\section{Participant Demographics}

Table~\ref{tab:participant_demographics} describes the demographic makeup of our 18-participant user study. The median participant age was 19.5 years (mean was 22.6, range 18--51 years). 

\label{sec:appendix_demographics}
\begin{table}[!h]
    \centering
    \small
    \caption{Demographic composition of the study sample ($N=18$).}
    \label{tab:participant_demographics}
    \begin{tabular}{@{}llrr@{}}
        \toprule
        \textbf{Dimension} & \textbf{Category} & \textbf{$n$} \\
        \midrule
        Age & 18--24 & 13 \\
          & 25--34 & 4 \\
          & 45--54 & 1 \\
        \midrule
        Gender & Male & 12 \\
             & Female & 6 \\
        \midrule
        Race/ethnicity & Asian & 11 \\
                     & White/Caucasian & 5 \\
                     & Middle Eastern & 1 \\
                     & Hispanic/Latino & 1 \\
        \midrule
        Education & Undergraduate & 14 \\
                & Graduate & 4 \\
        \bottomrule
    \end{tabular}
\end{table}

\









\end{document}